\documentclass[journal]{IEEEtran}
\usepackage{ifpdf}
\usepackage{threeparttable}

\usepackage{cite}
\ifCLASSINFOpdf
  \usepackage[pdftex]{graphicx}
  \usepackage{subcaption}
\else
\fi
\usepackage{todonotes}
\setuptodonotes{inline, color=green!20, size=\footnotesize}

\usepackage{amsmath}
\usepackage[siunitx]{circuitikz}
\usepackage{mathtools}
\usepackage{graphicx} % Required for inserting images
\usepackage{comment}
\usepackage{textcomp}
\usepackage{subcaption}
\usepackage{placeins}

\begin{document}

\title{Mixtenna: A Self-Biased Nonlinear Patch Antenna for Passive Third-Harmonic Radiation}

\author{Yishai~Brill,~\IEEEmembership{Member,~IEEE,}
        Yakir~Hadad,~\IEEEmembership{Senior Member,~IEEE,}
\thanks{The authors are with the School of Electrical Engineering, Tel-Aviv University, Tel Aviv 69978, Israel (corresponding author, Y.e-mail: yakirhadad@tauex.tau.ac.il).}}

% The paper headers
\markboth{October~2025}%
{Shell \MakeLowercase{\textit{et al.}}: Bare Demo of IEEEtran.cls for IEEE Journals}
\maketitle

\begin{abstract}
A nonlinear rectangular patch antenna (RPA) is presented in which back-to-back Schottky diodes are embedded at high-field regions to enable passive, bias-free harmonic generation. The self-biased diodes introduce a power-dependent impedance that drives efficient frequency up-conversion and selective third-harmonic radiation. A tailored matching network enhances third-harmonic excitation and coupling while preserving radiation efficiency at the fundamental frequency. Analytical modeling combined with SPICE-assisted full-wave time-domain simulations predicts strong odd-harmonic content, and measurements on RPA prototypes employing SMS7630 diodes confirm these results. Simulated and measured S-parameters and far-field patterns at 925 MHz and 2.775 GHz show excellent agreement. The demonstrated approach establishes nonlinear loading as an effective mechanism for passive harmonic control in compact radiators, enabling frequency-agile and spectrum-efficient antenna systems.
\end{abstract}

% Note that keywords are not normally used for peerreview papers.
\begin{IEEEkeywords}
Schottky diode, Rectangular Patch Antenna, Harmonic Radiation, Non-Linear Loading, Diode loaded antenna, Self biased diode.
\end{IEEEkeywords}

\IEEEpeerreviewmaketitle

\section{Introduction}

\IEEEPARstart{C}{onventional} electromagnetic wave engineering models antennas as linear time-invariant (LTI) systems. This assumption guarantees reciprocity between transmission and reception, simplifies cascaded channel modeling, and enables tractable analysis in both time and frequency domains. Relaxing the LTI assumption, however, unlocks a broad range of unconventional behaviors. 
%
%In a broader context, beyond antennas, recent demonstrations include passive nonlinear elements enabling broadband isolation~\cite{sounas2018broadband} and nonlinear circuit arrays exhibiting self-induced topological transitions~\cite{Hadad2018_NatElectron}.
%
The study of nonlinear antenna loading dates back several decades. Foundational work in the 1970s applied time-domain and Volterra-series techniques to wire antennas terminated with nonlinear loads~\cite{Schumann1974,Sarkar1976,LiuTesche1976,Kanda1980}. Subsequent efforts extended these concepts to arrays, more complex wire geometries, and harmonic nonlinear scattering~\cite{Gomez1987,HuangChu1993,Luebbers1993,Harger1976,Bucci1984}. Research in the 1980s–1990s further explored diode-loaded and active integrated microstrip antennas~\cite{Jenshan1994, ChangAIA2002, Navarro1991,Haskins1991,GuptaHall2000}, laying the foundations for modern nonlinear antenna modeling and design~\cite{Carvalho2002,Lee2000_Freq}.

A practical route to introducing nonlinearity in radiating systems is through the integration of nonlinear circuit elements, such as varactors or Schottky diodes, directly within the antenna structure. When externally biased, these components operate at a prescribed capacitance or resistance, enabling on–off switching or frequency tuning for small-signal RF control \cite{pozar, sounas2018broadband}. In contrast, when self-biased, their impedance becomes power dependent, and the operating point dynamically shifts with the amplitude of the incident RF signal~\cite{pozar,Adamski, Hadad2018_NatElectron}. Such self-biasing introduces an additional degree of freedom in antenna design, enabling functionalities that are unattainable with linear configurations.

Externally biased antennas have long been used to introduce controllable nonlinear or tunable behavior into radiating systems. By embedding varactors or Schottky diodes with appropriately designed bias networks, such antennas can achieve functions such as resonance tuning, on–off switching, dynamic filtering, polarization agility, and beam steering~\cite{deshmukh2018review,Haskins1991,Navarro1991}. These approaches extend the capabilities of fixed-geometry antennas but require auxiliary components—DC feeds, bias tees, RF chokes, and isolation networks—that complicate integration and become increasingly challenging to implement at high frequencies. Microstrip patch antennas, owing to their printed-circuit implementation and compatibility with surface-mount devices~\cite{Balanis,garg2001microstrip,james1989handbook,pozar_schaubert1995_microstrip}, have served as convenient platforms for such externally biased configurations, particularly in reconfigurable and active-integrated antenna concepts~\cite{GuptaHall2000,Jenshan1994,ChangAIA2002}.

Self-biased diodes, by contrast, operate passively and are extensively employed in RF energy-harvesting systems. When connected to antenna ports, they rectify incoming electromagnetic signals and exploit their intrinsic nonlinearity for RF-to-DC power conversion~\cite{sherazi2022comprehensive}. These rectifiers, commonly based on Schottky diodes, are typically optimized for the sub-6-GHz range, corresponding to the most abundant ambient energy sources such as cellular and Wi-Fi signals~\cite{fumtchum2019survey}.

Another established application of self-biased diodes is in harmonic passive RFID tags, which intentionally backscatter the incident reader signal at harmonic frequencies (e.g., $2f_0$, $3f_0$) to reduce interference and enhance detection in cluttered environments. Schottky diodes—with their low forward voltage and ability to conduct weak RF currents—enable compact, bias-free implementations~\cite{mondal2022recent}. Most reported harmonic tags emphasize second-harmonic generation, while only a few target the third harmonic due to the increased free-space path loss at higher frequencies, which limits the operational range. Measured conversion efficiencies for passive second-harmonic tags generally remain below --13\,dB ($\approx 5\%$)~\cite{palazzi2015lowpower,gu2018improving,presas2003high,colpitts1998harmonic}.

Several studies have also examined unintentional third-harmonic emissions from commercial UHF RFID tags, primarily addressing interference mitigation and the trade-off between signal-to-noise ratio (SNR) and receiver sensitivity~\cite{kumar2020harmonic,nikitin2009harmonic}. Because these analyses focus on spectral pollution rather than intentional harmonic enhancement, conversion efficiency is rarely quantified.

In this work, we investigate the nonlinear loading of rectangular microstrip patch antennas (RPAs) to achieve efficient third-harmonic radiation and to expand the electromagnetic functionality of compact antennas. A full-wave time-domain simulation framework incorporating SPICE-level nonlinear models is developed to capture the antenna’s dynamic behavior under realistic excitation conditions. The model allows precise control of excitation parameters—such as amplitude, frequency, and duty cycle—and enables the extraction of both reflection and radiation characteristics for frequency-domain analysis.
The proposed nonlinear RPA concept is validated through a combined simulation–measurement study. Special emphasis is placed on optimizing the coupling between the fundamental and third-harmonic modes while maintaining high radiation efficiency at the fundamental frequency. Fabricated prototypes confirm the predicted nonlinear behavior, demonstrating power-dependent third-harmonic radiation and nonlinear amplitude response. The results establish nonlinear loading as a viable method for achieving passive harmonic control, paving the way toward compact, frequency-agile, and spectrum-efficient antenna systems.

\section{Theoretical Modeling of a Diode as a Self-biased Nonlinear Load}
Considering a junction diode as the fundamental nonlinear element, the DC current–voltage relationship can be represented as a nonlinear current source, as expressed in~\eqref{eq:nl0}, 
\begin{equation}
I_D(V_D)=I_S(e^\frac{q V_D}{{kT}}-1)\label{eq:nl0}
\end{equation}
The I–V characteristic may be expanded into a power series about the operating point defined by the bias voltage $V_B$. Then, for a small RF harmonic signal, $\Delta v=v_0 \cos(\omega_0 t)$ (i.e., $V_D=V_B+\Delta v$, $v_0\ll V_B$) the current on the diode reads,   
\begin{align}
I_D(t) &= I_0 + I_1 \cos(\omega_0 t) + I_2 \cos(2\omega_0 t) + \notag \\
     &\quad +I_3 \cos(3\omega_0 t) + \cdots \label{eq:nl4}
\end{align}
%
%The current waveform in~\eqref{eq:nl4} illustrates the generation of higher-order harmonics. Here, $v_D$ denotes the diode voltage, while $\Delta v$ represents the RF excitation—a periodic signal at the fundamental angular frequency $\omega_0=2\pi f_0$ with amplitude $v_0$.

\begin{figure}[htbp]
    \centering
    \includegraphics[width=0.5\textwidth]{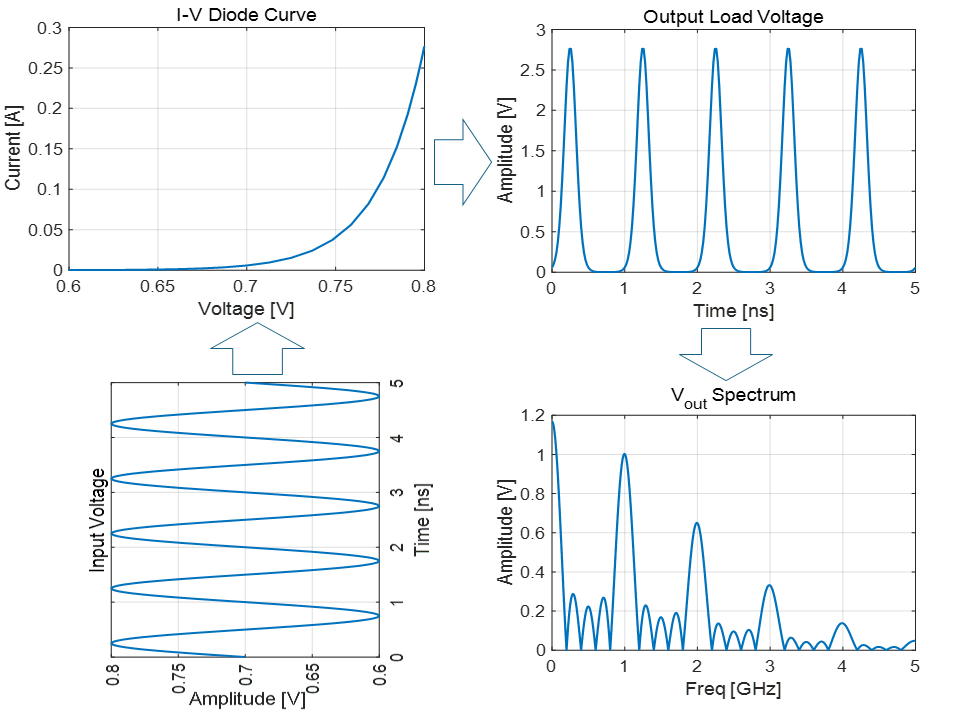}
    \caption{A graphic explanation showing a harmonic signal exciting a non linear device. The output signal is modified by the I-V curve relations and produces a harmonic-rich signal. }
    \label{fig:NL-load}
\end{figure}

In contrast to this small signal picture, as discussed in~\cite{pozar,Adamski}, when the diode is self-biased by the RF excitation and $V_B=0$ (i.e., no external DC bias), its operating point varies dynamically with the signal amplitude $v_0$. Consequently, self-biased diodes exhibit a power-dependent impedance and distinct behaviors under low- and high-power excitation. Figure~\ref{fig:NL-load} provides a graphical illustration of how a sinusoidal input applied to a nonlinear I–V characteristic produces an output rich in harmonic components.

%\begin{align}
%I_D(V_D)&=I_S(e^\frac{q V_D}{{kT}}-1)\label{eq:nl0}\\
%I_D(V_D) &= a_0 + a_1\Delta v+a_2\Delta v^2 +a_3\Delta v^3+...\label{eq:nl1} \\
%     V_D &= V_B + \Delta v\label{eq:nl2}\\
%     \Delta v &= v_0 \cos(\omega_0 t)\label{eq:nl3}\\
%     I_D(t) &= I_0 + I_1 \cos(\omega_0 t) + I_2 \cos(2\omega_0 t) + \notag \\
%&\quad +I_3 \cos(3\omega_0 t) + \cdots \label{eq:nl4}
% \end{align}

\subsection{Diode modeling under large signal excitation}
We extend the analytical modeling and describe the diode behavior under different conditions: forward and reverse bias, small and large input AC signals, and stored charge effects. As described in~\cite{antognetti1998semiconductor}, the static model of a real diode is given in equation~\eqref{eq:ID_fVD}. It divides the current into four distinct zones and takes into account the effects of large signal bias and breakdown.
\setlength{\arraycolsep}{2pt}
\renewcommand{\arraystretch}{1.1}

\begin{equation}
I_D(V_D)=
\begin{cases}
I_S\!\left(e^{\frac{V_D}{nV_T}}-1\right)+V_D\,\mathrm{GMIN}, & \text{ } V_D\ge -5nV_T,\\[4pt]
-\,I_S+V_D\,\mathrm{GMIN}, & \text{ }\mathclap{\ -BV< V_D<-5nV_T,}\\[4pt]
-\,I_{BV}, & \text{ } V_D=-BV,\\[4pt]
-\,I_S\!\left(e^{-\frac{BV+V_D}{V_T}}-1+\frac{BV}{V_T}\right), & \text{ } V_D<-BV.
\end{cases}
\label{eq:ID_fVD}
\end{equation}

To describe the response to an AC input signal, let us consider charge storage effects; otherwise, the device would be infinitely fast when switching from forward to reverse voltage. In a Schottky diode, the charge storage effects are dominated almost entirely by the junction capacitance equation~\eqref{eq:CD}:
\setlength{\arraycolsep}{2pt}
\renewcommand{\arraystretch}{1.1}
\begin{equation}
C_j(V_D) = 
\begin{cases}
{C_j(0) \left( 1 - \frac{V_D}{VJ} \right)^{-m}},
& \text{ } \mathclap{\ V_D < FC \times VJ} \\[4pt]
{\frac{C_j(0)}{A} \left( B + \frac{mV_D}{VJ} \right)},
& \text{ } \mathclap{\ V_D \geq FC \times VJ}\\[4pt]
\text{Where } A=(1-FC)^{(1+m)}\\[4pt] \text{ and } B=1-FC(1+m)\\[4pt]
\end{cases}\\
\label{eq:CD}
\end{equation}

Additionally, at high levels of bias, the current deviates from ideal behavior as a result of the ohmic resistance of the contacts as well as resistance associated with neutral regions in the device itself. These effects are modeled as a linear resistor $R_S$ in series with the diode.
The parameters used in equations~\eqref{eq:ID_fVD} and ~\eqref{eq:CD} are the formal keywords that appear in the SPICE .MODEL statement.
Other parameters are: $T=300$K, $q=1.602e-19$C, $k=1.38e-23$J/K and ${\rm GMIN}=1e-12$mho.\\
Table~\ref{tab:sms7630_spice} describes the SPICE parameters and their physical meaning and values for the SMS7630 Schottky diode~\cite{skyworks2}.

\subsection{Schottky Diode}
A Schottky diode is a type of semiconductor diode formed by a junction of a metal with a semiconductor. It has a short transit time due to the absence of minority charge carrier storage. These unique properties make it ideal for very fast switching action and high-frequency applications.
It also has a low forward voltage drop; therefore, self-biasing non-linear effects may be obtained with relatively low RF power.\\
In this work, we chose the SMS7630-079LF surface mount Schottky diode from SKYWORKS corporation \cite{skyworks2}. It is a single silicon junction in a SC-79 plastic package and has a typical forward voltage of 0.135 to 0.24 volts and a transit time of 10pS.

\subsection{Back-to-Back diode configuration (Clipper)}
In this work, the non-linear load of interest is a dual-diode load. Two diodes are placed in a back-to-Back configuration in order to generate odd harmonics of the fundamental excitation frequency. This setup is known as a Clipper circuit~\cite{skyworks1}. It allows low amplitude input signals to pass unchanged to the output port. When the amplitude reaches the limiting voltage, the diodes are forced into conduction, and the output signal is "clipped" to this threshold value. Figure~\ref{fig:ClipperSchematic} and Figure~\ref{fig:LargeSignalSchottkyDiodeModel} depicts the clipper circuit scheme and the large signal diode model with its non-linear current source $I_D$, voltage dependent junction capacitance $C_j(V_j)$ and the resistor $R_S$.\\
To understand this mechanism, we will analyze it in the time domain and then move to the frequency domain to demonstrate the generation of odd, and especially third order harmonics.

\subsection{Time domain analysis}
Writing the KCL equations for the circuit in Figure~\ref{fig:ClipperSchematic} :
\begin{align}
     I_{\rm total}&=\frac{V_s(t)-v}{R_{src}}=i_1+i_2
\end{align}
Every diode branch of the clipper circuit in Figure~\ref{fig:ClipperSchematic} is analyzed as its large signal equivalent model shown in Figure~\ref{fig:LargeSignalSchottkyDiodeModel}:
\begin{align}
    &\text{Branch D1: } \nonumber\\
    i_1 &= i_D(v_1)+C_{j1}(v_1)\frac{dv_1}{dt}\\
    v_1 &=v-i_1R_{S1}\\
    &\text{Branch D2, flipped diode: }\nonumber\\
     i_2 &= -i_D(-v_2)+C_{j2}(-v_2)\frac{dv_2}{dt}\\
    v_2 &=v-i_2R_{S2}
\end{align}

At every time step, we are solving this non-linear set of equations. Since the current depends on its instantaneous voltage and its derivative, which itself depends on previous voltage, it is implicit in time. This is solved through a time stepping scheme with a relaxed update iteration.

The analytical clipper circuit model is implemented in MATLAB. The total current results, $I_{total}$, are compared to a commercial SPICE engine simulation (LTspice, ~\cite{LTspice}). The simulation inputs are: $R_{\rm src}=50\Omega$, $R_{S1}=R_{S2}=20\Omega$. The excitation is a sinusoidal signal, $V_s(t)=A_0\sin(2\pi f_0t)$, $f_0=925MHz$, and $A_0$ are set for input power levels of -10dBm and -3dBm. Figure~\ref{fig:TimeDomainResults} shows the excellent agreement between the analytical model and the results of the spice simulation based on the SMS7630 diode model. Figure~\ref{fig:FFTResults} shows how only odd harmonics are generated in the signal spectrum, while canceling out the DC and even terms. Appendix~\ref{appendix: Clipper_freq_domain_analysis} is a complementary analysis of the clipper circuit in the frequency domain.

\section{Simulations and testing of the clipper load}
\subsection{The Approach}
Our next step is to integrate the anti-parallel diode pair into the 3D electromagnetic solver, CST Studio~\cite{CST}. This software package includes a SPICE engine that is optimized for interaction with the electromagnetic domain. Its time-domain solver supports the inclusion of component-level devices when they are described as SPICE circuit files.

\begin{figure}[!t]
\centering
\begin{subfigure}{0.67\columnwidth}
  \centering
  \resizebox{\linewidth}{!}{%
  \begin{circuitikz}[straight voltages][scale=1, transform shape,
    every node/.style={font=\normalsize}]
    \draw
      (0,0) node[ground] {} to[vsourcesin, l_=$V_s(t)$] (0,4)
      to[R, l_=$R_{\mathrm{src}}$] (2,4)
      to[short, i_=$I_{\mathrm{total}}$] (4,4)
      to[short, i=$i_1$] (4,2.5)
      to[full diode, l=$D_1$] (4,1.5) -- (4,0)
      (4,0) -- (0,0)
      (7,4) to[short, i=$i_2$] (7,2.5)
      (4,0) -- (7,0) to[full diode, l=$D_2$] (7,4) -- (4,4)
      (3,0) to[open, v^>=$v$] (3,4);
  \end{circuitikz}}
  \caption{}
  \label{fig:ClipperSchematic}
\end{subfigure}\hspace{0 pt}
\begin{subfigure}{0.31\columnwidth}
  \centering
  \resizebox{\linewidth}{!}{%
  \begin{circuitikz}[scale=1, transform shape,
    every node/.style={font=\normalsize}]
    \draw
      (0,0) to[short, o-] (0,1) -- (-1,1)
      to[american current source, l_=$I_{D}$, invert] (-1,4) -- (0,4)
      to[R, l=$R_s$, -o] (0,6)
      (0,4) -- (1,4)
      to[variable capacitor, l=$C_j(V_j)$, invert] (1,1) -- (0,1);
  \end{circuitikz}}
  \caption{}
  \label{fig:LargeSignalSchottkyDiodeModel}
\end{subfigure}
\caption{(a) Anti-parallel diode pair connected to a sinusoidal source with internal resistance $R_{\mathrm{src}}$. (b) Schottky diode large-signal equivalent model.}
\label{fig:ClipperSchematicMain}
\end{figure}
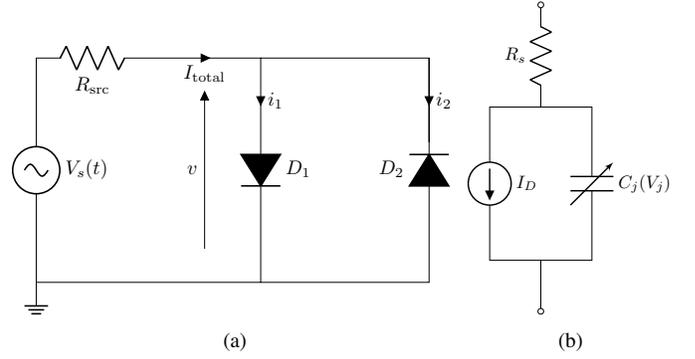

\begin{figure}[!t]
    \centering
    % --- Left image ---
    \begin{subfigure}[b]{0.97\columnwidth}
        \centering
        \includegraphics[width=\linewidth]{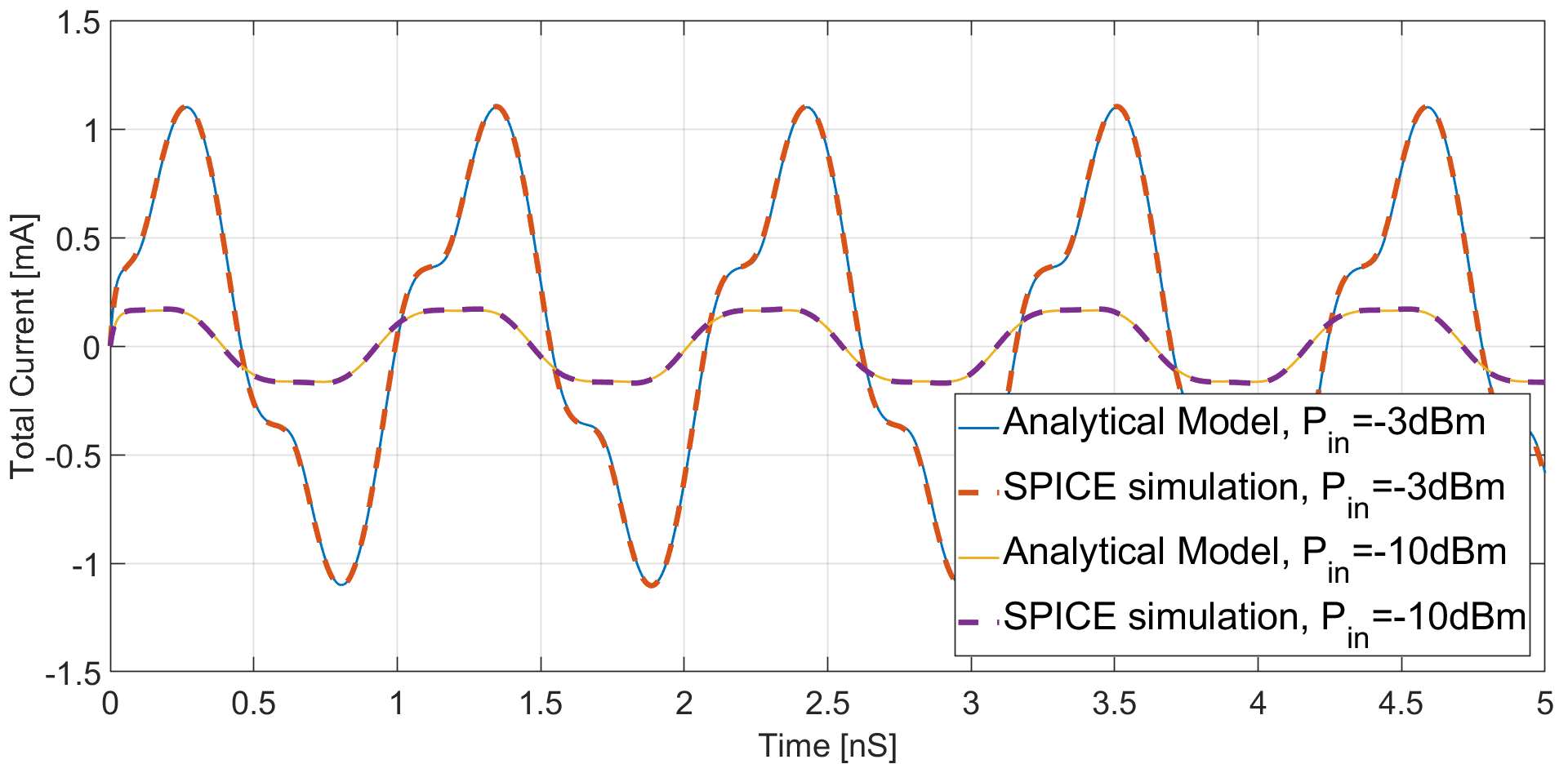}
        \caption{}
        \label{fig:TimeDomainResults}
    \end{subfigure}
    \hfill
    % --- Right image ---
    \begin{subfigure}[b]{0.97\columnwidth}
        \centering
        \includegraphics[width=\linewidth]{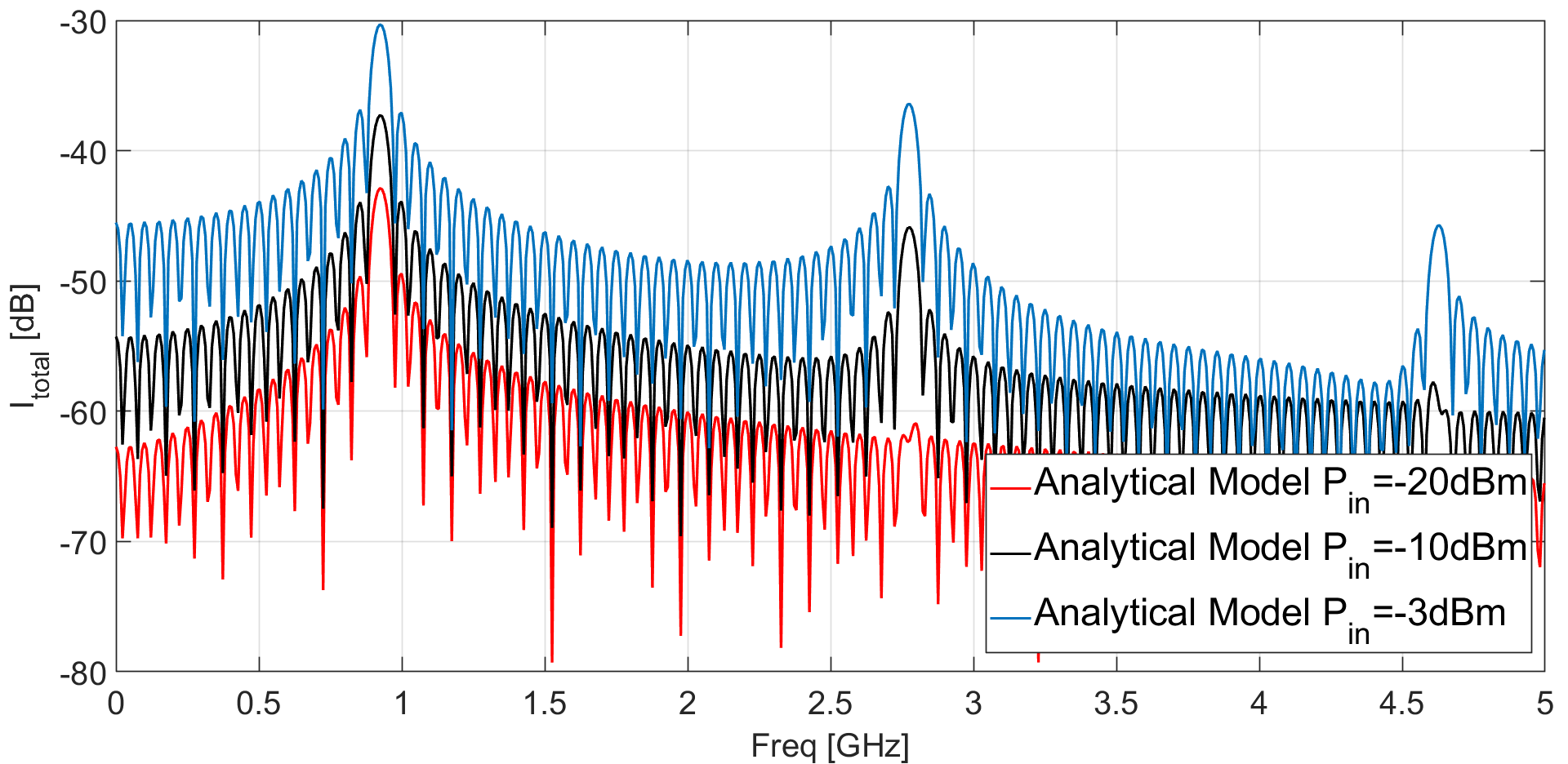}
        \caption{}
        \label{fig:FFTResults}
    \end{subfigure}

    \caption{(a) Comparison between analytical model and SPICE simulation of the clipper circuit. We compare the total current, $I_{\text{total}}$, results for input powers of -10~dBm and -3~dBm.
    (b) FFT results of the total current signals from the analytical model simulation. The fundamental frequency, $f_0 = 925$~MHz, and its third harmonic at $f_3 = 2775$~MHz are clearly visible. The second harmonic is completely canceled.}
    \label{fig:ClipperResultsCombined}
\end{figure}

Our simulation strategy is to establish a transient environment in which a user-defined excitation waveform is applied. The waveform parameters—frequency, amplitude, T\textsubscript{on} ,T\textsubscript{off} and T\textsubscript{total} are controlled through custom code. The waveform is injected into the 3D model, and the solver captures the propagating, conducted, and radiated signals for subsequent post-processing. Additional macros are implemented to automate solver execution and to generate performance metrics that are not directly available from the standard simulation output.\\
Several studies have reported experimental and analytical techniques for extracting Schottky diode parameters over wide frequency ranges and under both small- and large-signal conditions. In~\cite{chen2020schottky}, the authors employ a commercial bench-top semiconductor analyzer to measure the junction capacitance under various bias conditions, followed by S-parameter characterization on a dedicated PCB. The results are incorporated into an ADS model for a specific diode package, providing both small- and large-signal representations.
In~\cite{tang2013analytical}, analytical formulations for junction parameters are combined with on-wafer broadband S-parameter measurements and detailed 3D electromagnetic simulations. This hybrid approach enables reliable parameter extraction, including parasitic effects, under small-signal operation.

In this work, we use vendor-supplied SPICE parameters that include package parasitic capacitance and inductance~\cite{skyworks2} and validate the results through clipper-PCB measurements. Future work will apply parameter extraction methodologies that are similar to the ones described in~\cite{chen2020schottky} and~\cite{tang2013analytical} to further improve model accuracy.

\subsection{Simulations and Measurements}
\begin{figure}[!t]
    \centering
    % --- Left image ---
    \begin{subfigure}[b]{0.48\columnwidth}
        \centering
        \includegraphics[width=\linewidth]{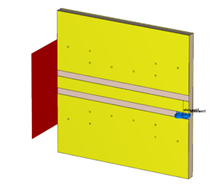}
        \caption{}
        \label{fig:clipper_pcb_sim}
    \end{subfigure}
    \hfill
    % --- Right image ---
    \begin{subfigure}[b]{0.48\columnwidth}
        \centering
        \includegraphics[width=\linewidth]{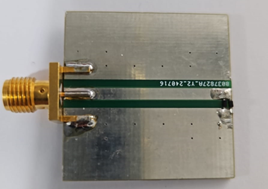}
        \caption{}
        \label{fig:clipper_pcb_lab}
    \end{subfigure}
\hfill
 \begin{subfigure}[b]{0.97\columnwidth}
        \centering
        \includegraphics[width=\linewidth]{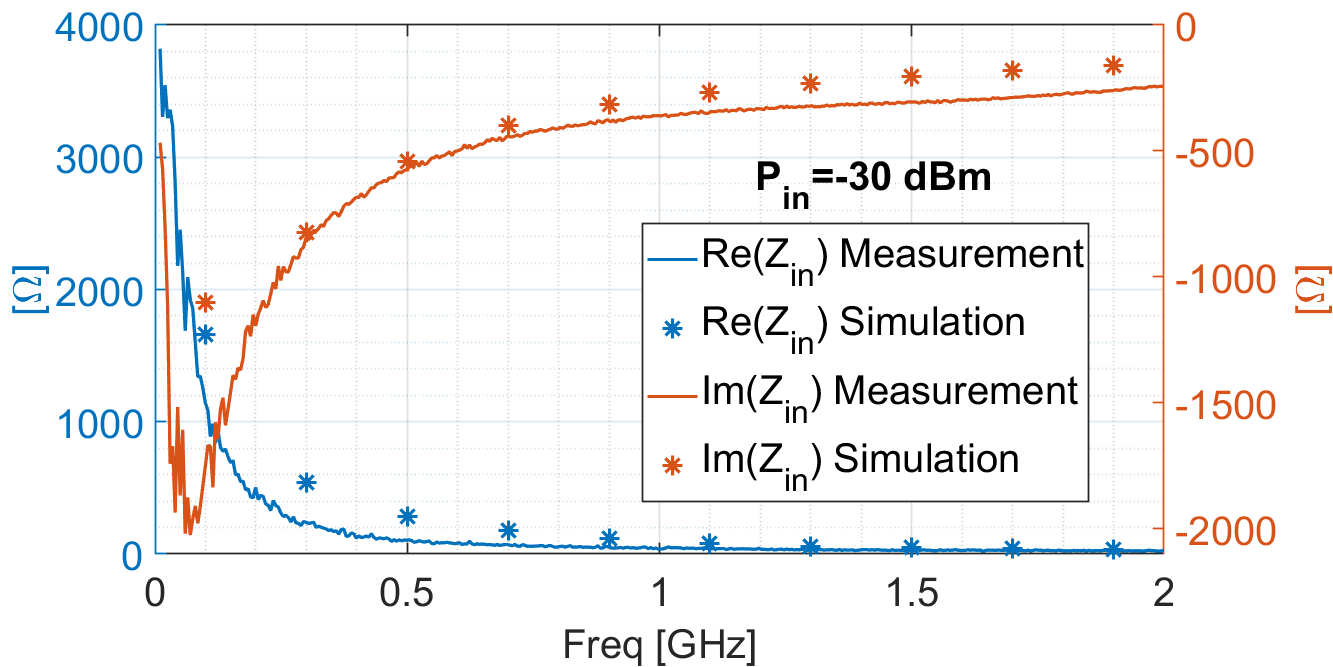}
        \caption{}
        \label{fig:ClipperImpedanceSmallSignal}
    \end{subfigure}
    \hfill
    \begin{subfigure}[b]{0.97\columnwidth}
        \includegraphics[width=\linewidth]{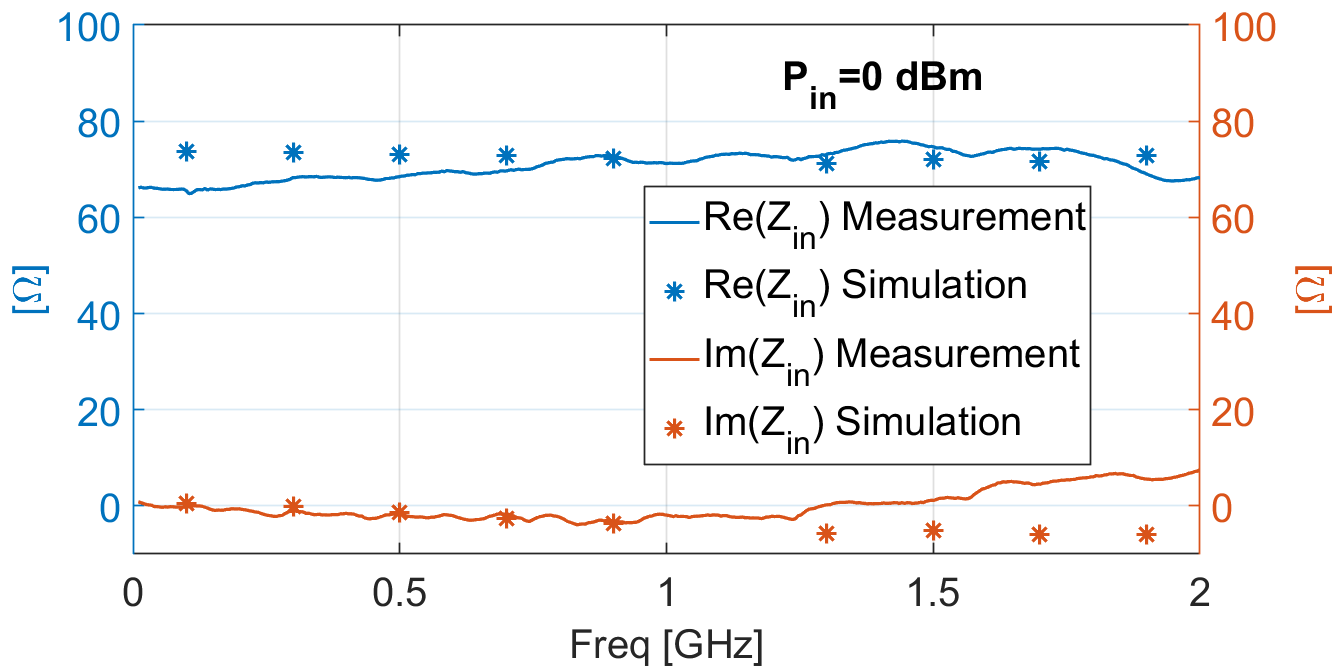}
        \caption{}
        \label{fig:ClipperImpedanceLargeSignal}
    \end{subfigure}
    \caption{(a) Simulation model of the PCB coupon used for diode impedance characterization. 
    (b) Fabricated PCB coupon used for diode impedance characterization. 
    The SMS7630 back-to-back diode pair is soldered on the right open-ended edge.\\
    (c) Are the Real and Imaginary part of Z\textsubscript{in} parts results vs. frequency for P\textsubscript{in}=-30dBm (small signal condition). (d) Are the Real and Imaginary part of Z\textsubscript{in} parts results vs. frequency for P\textsubscript{in}=0dBm (large signal condition).}
    \label{fig:clipper_pcb_comparison}
\end{figure}

 A back-to-back diode pair, i.e. clipper, was soldered to a test PCB at the edge of an open microstrip line in shunt to the ground, as seen on Figure~\ref{fig:clipper_pcb_lab}. An identical CAD, depicted in Figure~\ref{fig:clipper_pcb_sim}, was modeled in the EM simulator along with the corresponding SPICE files. A CW signal is injected to the port and the input impedance is calculated from the reflection coefficient data:

\begin{equation}
\tilde{Z}_{\text{in}} = Z_0 \cdot \left( \frac{1 + \tilde{\Gamma}}{1 - \tilde{\Gamma}} \right)
\label{eq:inputIm}
\end{equation}

The frequency is swept in the simulation from 100 MHz to 2100 MHz and the CW signal power ranges from -30 dBm to +10 dBm.
The reflection coefficient was measured with a VNA following calibration and a reference plane shift to the open end of the test PCB. The input impedance was calculated as in equation~\eqref{eq:inputIm}.

The self-biased Schottky clipper simulations agree well with the measurements. We have obtained its complex impedance values over a wide frequency and amplitude range. In Figure~\ref{fig:ClipperImpedanceSmallSignal} we can see that under small signal input, the clipper is mainly capacitive with $Z_{in}=44.2-j380.1[\Omega]$ at $f_0=925$MHz and $P_{\rm in}=-30$dBm. In Figure~\ref{fig:ClipperImpedanceLargeSignal}, under large signal input, the clipper is mainly resistive with $Z_{\rm in}=71.8-j3.4[\Omega]$ at $f_0=925$MHz and $P_{\rm in}=0$dBm. The SPICE parameters that were used for each single diode in the simulations are summarized in Table~\ref{tab:sms7630_spice}.

\begin{table}[htbp]
\caption{SPICE Parameters for SMS7630 Diode}
\label{tab:sms7630_spice}
\centering
\begin{tabular}{{lccc}}
\hline
\textbf{Parameter} & \textbf{Value} & \textbf{Unit} & \textbf{Description} \\
\hline
L & 7e-10 & H & Parasitic series inductance (package) \\
C & 7e-14 & F & Parasitic parallel capacitance (package) \\
IS & 5e-06 & A & Saturation current ($I_s$) \\
RS & 30.0 & Ohm & Ohmic resistance ($R_S$)\\
N & 1.05 & - & Emission coefficient ($n$) \\
EG & 0.69 & eV & Energy gap \\
XTI & 2.0 & - & Saturation current temperature exponent \\
BV & 2.0 & V & Reverse breakdown voltage \\
IBV & 0.0001 & A & Reverse breakdown current ($I_{BV})$ \\
CJO & 1.4e-13 & F & Zero bias junction capacitance ($C_j(0)$)\\
VJ & 0.34 & V & Junction potential \\
M & 0.4 & - & Grading coefficient ($m$)\\
FC & 0.5 & - & Coefficient for forward \\
      &       &    &bias depletion capacitance\\
TT & 1e-11 & s & Transit time \\
KF & 0.0 & - & Flicker noise coefficient \\
AF & 1.0 & - & Flicker noise exponent \\
\hline
\end{tabular}
\end{table}

 \section{Antenna baseline}
 \subsection{Rectangular patch antenna (RPA)}
 After establishing a validated clipper model through simulations, the next step is to integrate the nonlinear load with an antenna. The process begins with the design of a rectangular patch antenna (RPA) as a baseline, followed by optimization for nonlinear loading and efficient harmonic generation.

A commonly accepted configuration for this antenna is a rectangular patch printed on a dielectric substrate and edge-fed at its center by a microstrip line, as described in~\cite{Balanis}. The patch length, typically on the order of  $\lambda/ 2\sqrt{\varepsilon_r}$, excites the fundamental TM\textsubscript{01} mode of the structure. This design generally provides a matched impedance bandwidth of about 2\% and a directive radiation pattern, with total radiation efficiency ranging from 30\% to 70\%, depending on the dielectric losses.

\begin{figure}[!t]
    \centering
    % --- First image ---
    \begin{subfigure}{\linewidth}
        \centering
        \includegraphics[width=0.9\linewidth]{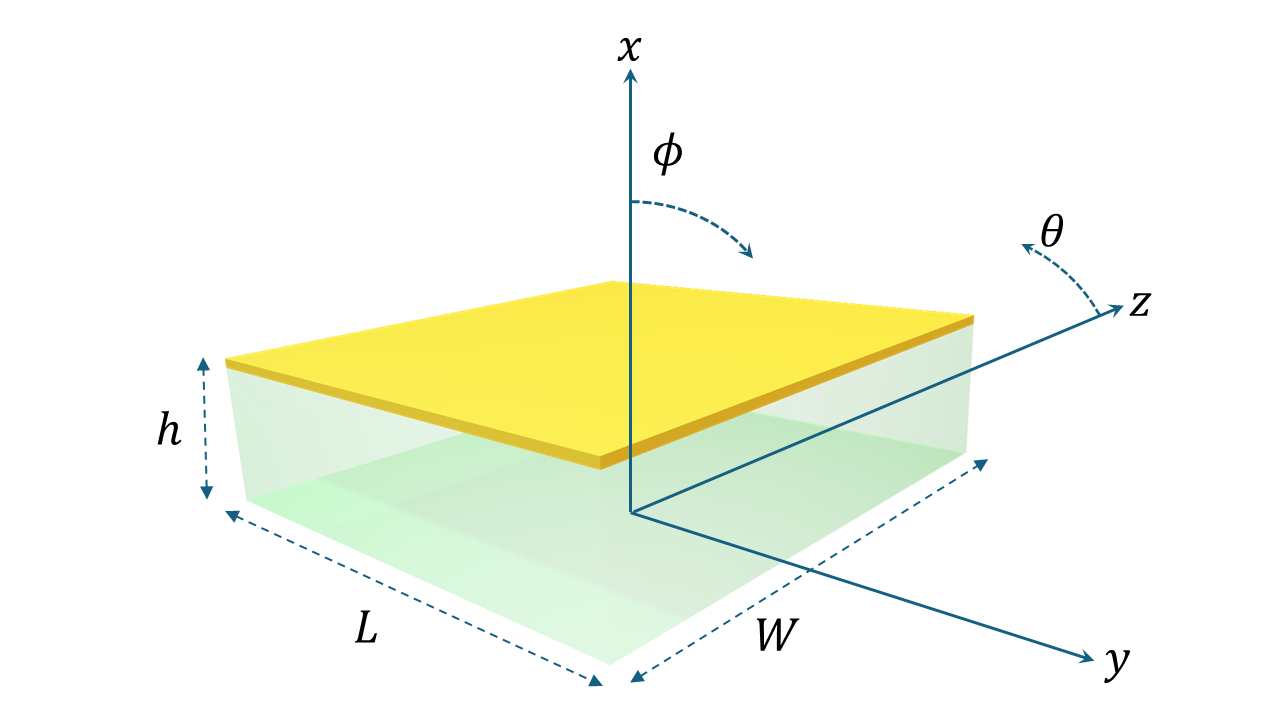}
        \caption{}
        \label{fig:cavity_model}
    \end{subfigure}

    \vspace{0.8em} % controls vertical spacing between images

    % --- Second image ---
    \begin{subfigure}{\linewidth}
        \centering
        \includegraphics[width=0.75\linewidth]{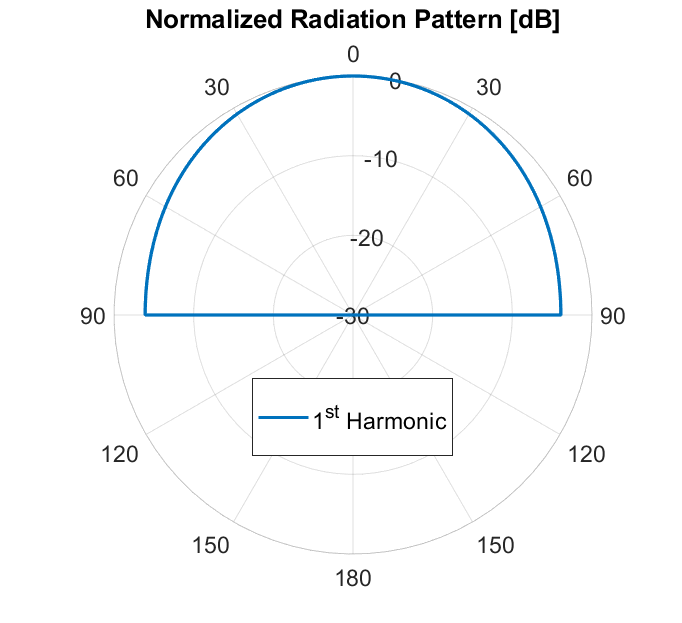}
        \caption{}
        \label{fig:theoretical_pat_1st_harmonic}
    \end{subfigure}

    \caption{(a) The cavity model and its parameters. (b) RPA theoretical normalized radiation pattern in the XY plane of figure~\ref{fig:cavity_model}. The RPA parameters are: W=113.25mm, L=93.5mm, h=1.52mm, f\textsubscript{0}=925 MHz. }
    \label{fig:cavity_and_theoretical_pattern}
\end{figure}

The radiation pattern of the RPA is determined by the modes that propagate within the patch at the frequency of interest. For the first-harmonic radiation pattern, only the TM\textsubscript{01} mode is considered. Using the cavity model presented in~\cite[Ch.~14]{Balanis}, the radiation pattern can be described as the combined field of two magnetic dipoles. The cavity geometry and coordinate system are shown in Fig.~\ref{fig:cavity_model}.

\begin{align}
\label{eq:rpa_pattern}
E_{\phi} &= \frac{+j k_0 h W E_0 e^{-j k_0 r}}{\pi r} 
\left\{ \sin\theta \cdot \frac{\sin X}{X} \cdot \frac{\sin Z}{Z} \right\}\cdot \\
&\quad  \cdot \cos\left( \frac{k_0 L_{\text{eff}}}{2} \sin\theta \sin\phi\right) \notag;\\
&\quad  X=\frac{k_0h}{2} \sin\theta \cos \phi; Z=\frac{k_0W}{2} \cos \phi\notag 
\end{align}

Where, $k_0$ is the free space wavelength, $E_0$ is the excitation amplitude, $r$ is the distance of field evaluation and $L_{\rm eff}$ is the effective patch length due to the fringing field effect. In our case, $L$=93.5mm and $L_{\rm eff}$=95mm which is 1.6\% longer due to the fringing field effect.

At the 3\textsuperscript{rd} harmonic, the effective electrical spacing between the equivalent magnetic dipoles increases, potentially risking grating lobes. However, using the criterion \cite{Balanis} $L<\lambda/(1+|\cos\theta_0|)$ with $\theta_0=90^\circ$, we find $L_{\rm max}<\lambda$. In our design, $L=93.5$mm and $\lambda\approx108$mm at $2.775$GHz, so grating lobes are still avoided. %as confirmed by the radiation patterns in Fig.~\ref{fig:theoretical_pat_1st_harmonic}.  

%In the 3\textsuperscript{rd} harmonic frequency the electrical length between the two magnetic dipoles is $3\lambda/ 2\sqrt{\varepsilon_r}$. Therefore, grating lobes may appear in front half-space direction. As described in \cite{Balanis} ch.6, the maximal distance between two radiating elements before grating lobes appear in the horizon is:
%\begin{align}
%    L_{max}<\frac{\lambda}{1+|\cos\theta_0|}\xrightarrow{\theta_0=90^0} L_{max}<\lambda
%\end{align}

%In our case, L\textsubscript{max}=93.5mm and $\lambda$ at f=2775 MHz is 108mm. The two-slot array equivalence shows that in the third harmonic frequency grating lobes are still avoided, and a single dominant maxima in the radiation pattern is obtained in the front half-space. Figure~\ref{fig:theoretical_pat_1st_harmonic} shows the calculated radiation patterns in the first harmonic according to equation~\eqref{eq:rpa_pattern} .

\subsection{Simulation results of the RPA}
\begin{figure}[!t]
    \centering
    % --- Left image ---
    \begin{subfigure}[b]{0.48\columnwidth}
        \centering
        \includegraphics[width=\linewidth]{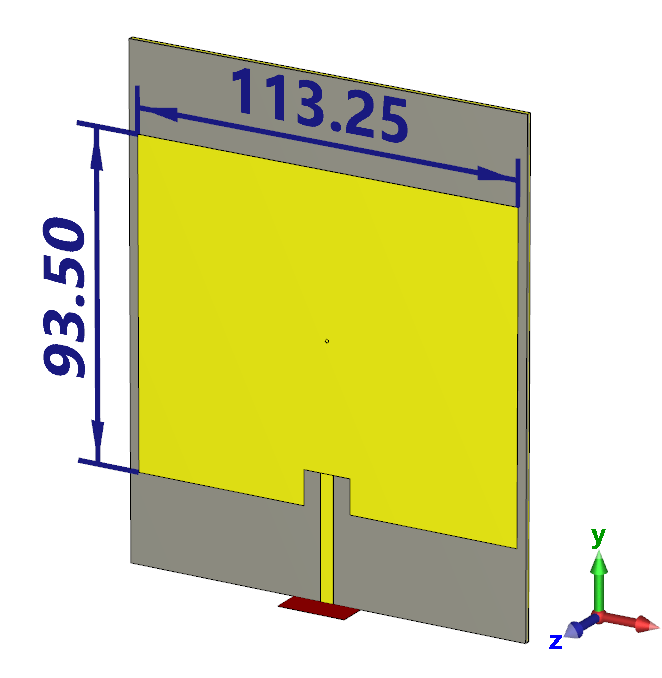}
        \caption{}
        \label{fig:Teflon_rpa}
    \end{subfigure}
    \hfill
    % --- Right image ---
    \begin{subfigure}[b]{0.5\columnwidth}
        \centering
        \includegraphics[width=\linewidth]{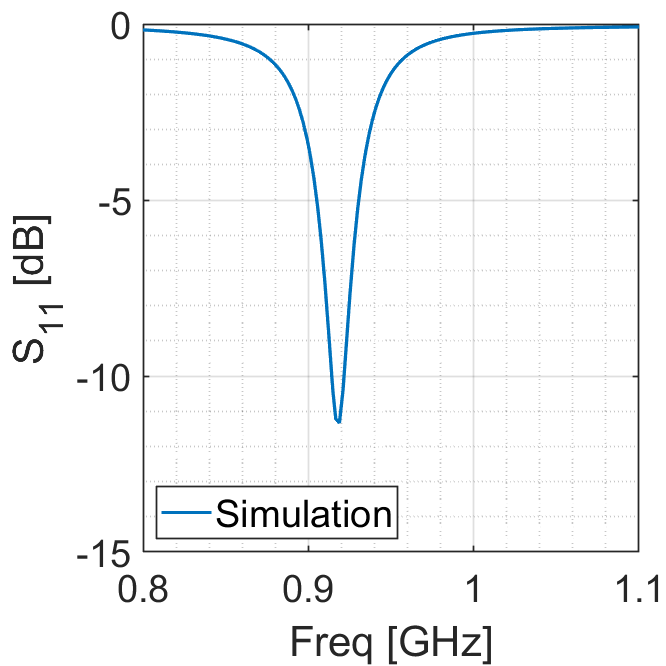}
        \caption{}
        \label{fig:rpa_teflon_s11}
    \end{subfigure}

    \caption{(a) The Teflon substrate RPA antenna model. Dimensions are in millimeters. (b) Simulation results of the Teflon RPA antenna's S\textsubscript{11} vs. frequency. }
    \label{fig:clipper_pcb_and_impedance}
\end{figure}

The performance of the RPA was simulated on two dielectric substrates: FR-4 and Teflon, each with a thickness of 60 mil (1.52 mm). An inset feed was implemented to achieve optimal impedance matching to a $50\Omega$ system. The antenna and substrate dimensions were optimized for efficient operation at 915 MHz, and the results are summarized in Table~\ref{tab:rpa_performance}.
Full-wave 3D simulations of the RPA yielded peak directivities of 6.3 dBi and 5 dBi at 925 MHz and 2775 MHz, respectively. As shown in Fig.~\ref{fig:Teflon_rpa}, the Teflon-based RPA exhibits higher radiation efficiency than the FR-4 design and is therefore selected as the baseline for subsequent analysis.

\begin{table}[htbp]
\caption{Performance comparison of two RPAs on different substrates}
\label{tab:rpa_performance}
\centering
\begin{tabular}{lcc}
\hline
\textbf{Parameter} & \textbf{FR-4} & \textbf{Teflon} \\
\hline
Dielectric constant & 4.3 & 3.0 \\
Loss tangent        & 0.025 & 0.0018 \\
Radiation efficiency (\%) & 33 & 81 \\
Total radiation efficiency\newline (including matching) (\%) & 32 & 73 \\
\hline
\end{tabular}
\end{table}

\section{Patch Antenna under Non-linear loading}
This part focuses on odd harmonic radiation and frequency conversion. Specifically, optimization and enhancement of 1\textsuperscript{st}-to-3\textsuperscript{rd} frequency conversion and radiation efficiency.\\
We now introduce a clipper model to the baseline design of the RPA as shown in Figure~\ref{fig:userdef-cw-and-box}. The clipper model is based on the vendor's SPICE description. Embedding the clipper in high E-Field areas of the RPA and referencing it to the antenna's ground will generate odd-order current components on the antenna's conductors, as noted in equation~\eqref{eq:SumOddTermsOnly} of Appendix A.\\
High E-Field areas are dictated by the modes that propagate in the RPA. The  fundamental radiated mode is TM\textsubscript{01} and maximal E-Field is located at the bottom and top edges of the patch. When analyzing the modes in the 100 MHz to 3 GHz frequency range, we see many potential modes that can propagate in the baseline RPA geometry. Due to the clipper loading, we are particularly interested in the 1:3 mode frequency ratio. When viewing the field pattern at 3 times the initial TM\textsubscript{01} mode frequency (i.e. 3\textsuperscript{rd} harmonic), we see that TM\textsubscript{03} also exists in this geometry. The top and bottom RPA edges are common areas for maximal E-field for both TM\textsubscript{01} and TM\textsubscript{03} modes. Therefore, the clipper is embedded at the center of the top RPA edge in shunt to the patch radiator and referenced to the ground.

\subsection{Simulation environment}
The RPA model is enclosed within an air box defined by radiation boundary conditions, and an E-field probe is positioned in the far field at a distance corresponding to the third-harmonic wavelength. A continuous-wave (CW) sinusoidal signal is applied at the microstrip port with a specified frequency and amplitude to excite the antenna.

\begin{figure}[htbp]
    \centering
    \includegraphics[width=\linewidth]{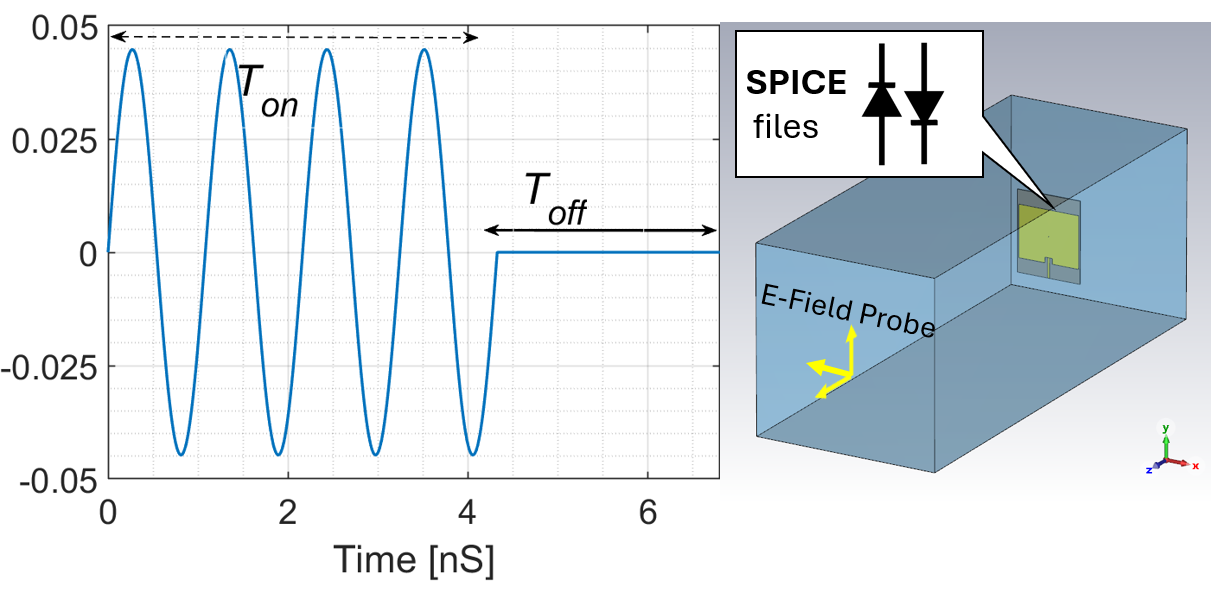}
    \caption{Right: Simulation environment for the RPA model. The RPA is in a radiation box and an E-field probe in its far-field region. Left: An excitation CW signal set for f\textsubscript{0}=925MHz and 0dBm power level.}
    \label{fig:userdef-cw-and-box}
\end{figure}

\begin{figure}[!t]
    \centering
    % --- First image ---
    \begin{subfigure}{\linewidth}
        \centering
        \includegraphics[width=0.97\linewidth]{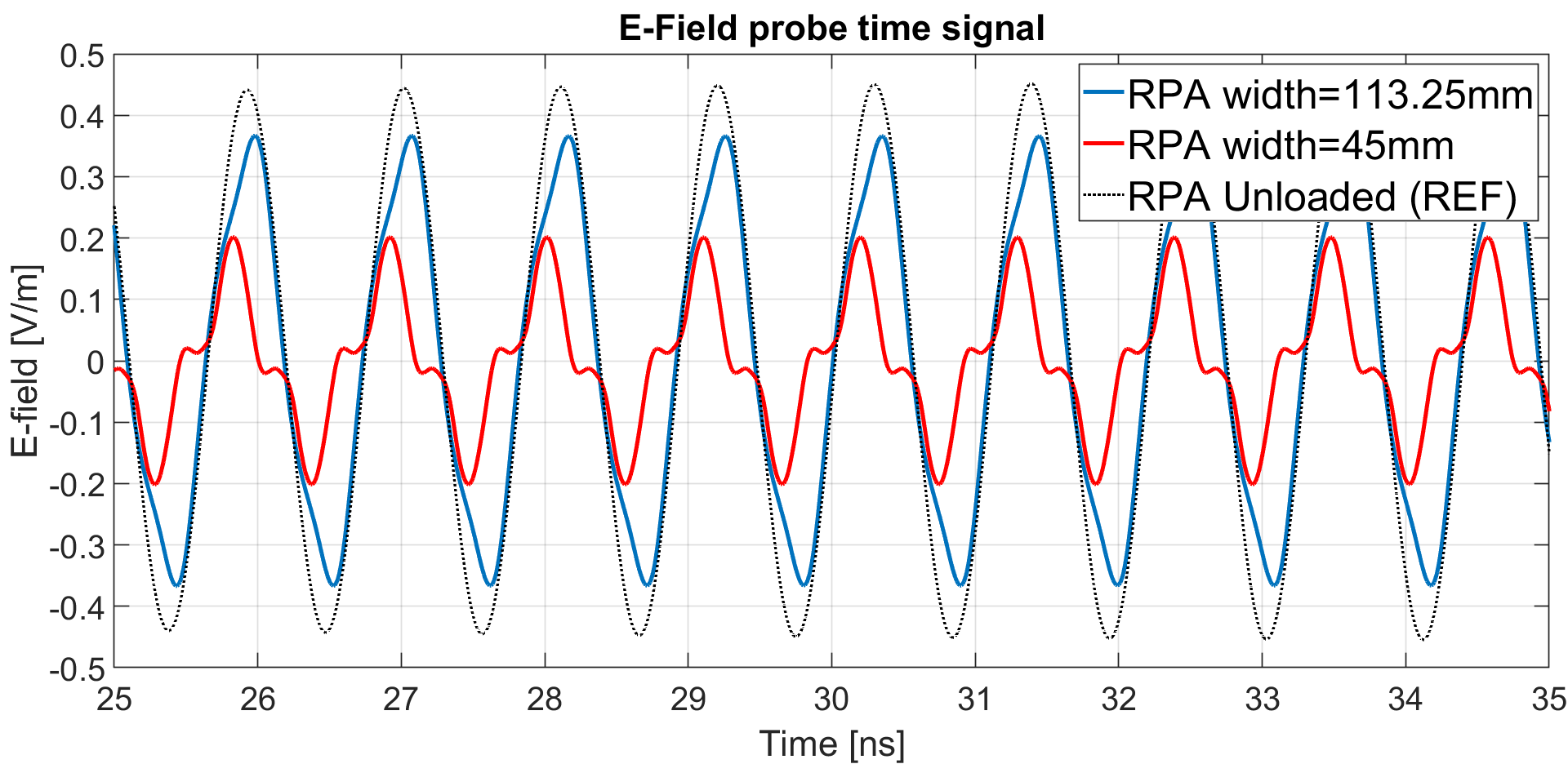}
        \caption{}
        \label{fig:rpa-time-sig}
    \end{subfigure}

    \vspace{0.8em} % controls vertical spacing between images

    % --- Second image ---
    \begin{subfigure}{\linewidth}
        \centering
        \includegraphics[width=0.97\linewidth]{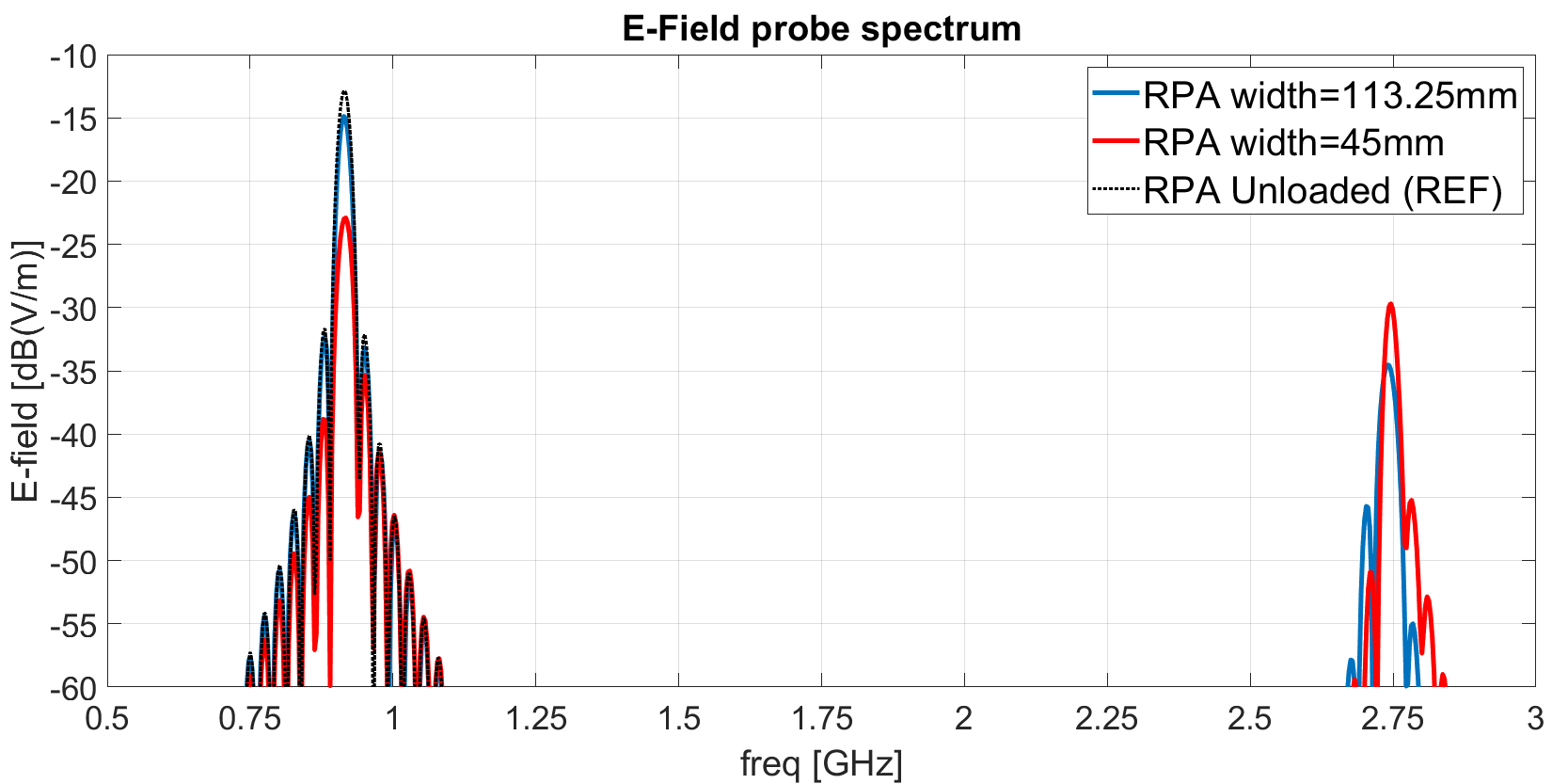}
        \caption{}
        \label{fig:rpa-freq-dom}
    \end{subfigure}

    \caption{(a) E-field probe received signals in the time domain simulation for different RPA configurations: Unloaded antenna, clipper loaded RPA and patch width=113.25mm, clipper loaded RPA and patch width=45mm. (b) E-field probe frequency responses for the corresponding RPA configurations.}
    \label{fig:rpa-freq-and-time-results}
\end{figure}

The E-field probe records the transient electric field as a function of time. The recorded signal is then transformed into the frequency domain using a Fast Fourier Transform (FFT), and the harmonic peak values are extracted for analysis. Figure~\ref{fig:userdef-cw-and-box} illustrates the simulation setup for the RPA model and the CW excitation signal. The E-field probe captures the transient transmitted field, with the resulting time-domain responses shown in Fig.~\ref{fig:rpa-time-sig} and their corresponding frequency-domain spectra presented in Fig.~\ref{fig:rpa-freq-dom}.

\subsection{Modification in the RPA width}
The RPA width was gradually reduced to optimize the radiation intensity of the 1\textsuperscript{st} and 3\textsuperscript{rd} harmonics in the far-field probe spectrum. The initial width of 113.25 mm was optimized to 45 mm based on simulation results.

\begin{figure}[htbp]
    \centering
    \includegraphics[width=0.9\linewidth]{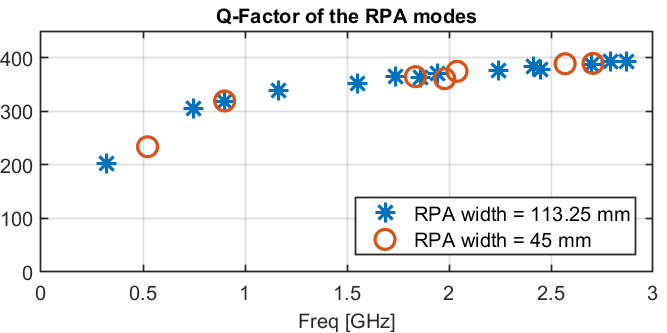}
    \caption{Q-Factor of the modes in the wide and narrow RPA geometries}
    \label{fig:rpa-modes}
\end{figure}

\begin{figure}[!t]
    \centering
    \begin{subfigure}[b]{0.6\columnwidth}
        \centering
        \includegraphics[width=\linewidth]{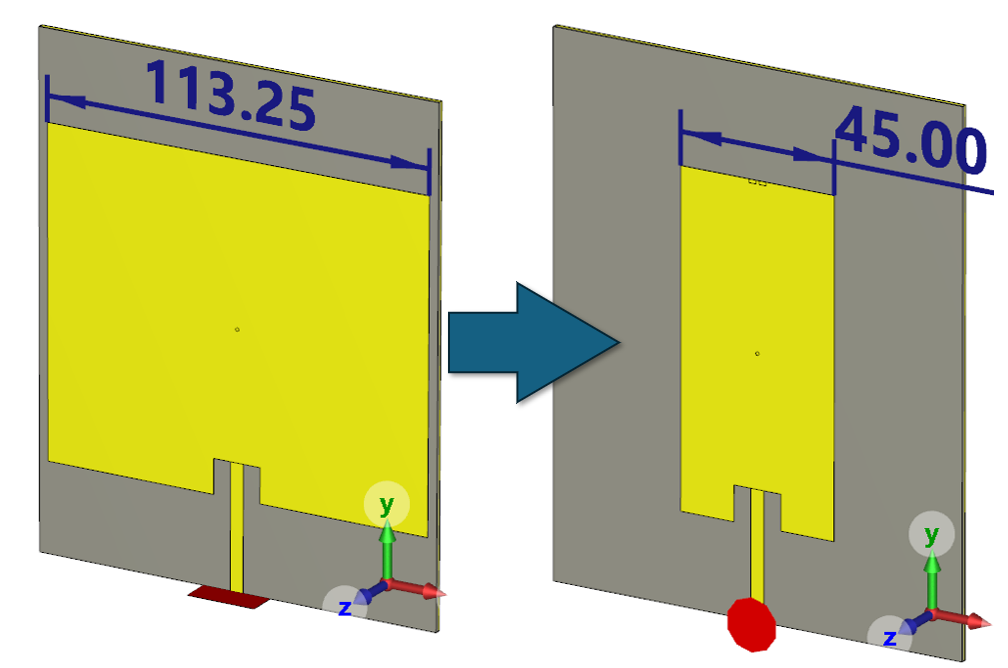}
        \caption{}
        \label{fig:rpa_width_reduction}
    \end{subfigure}
    \hfill
    \begin{subfigure}[b]{0.37\columnwidth}
        \includegraphics[width=\linewidth]{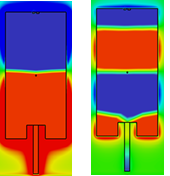}
        \caption{}
        \label{fig:rpa_maintained_modes}
    \end{subfigure}
    \begin{subfigure}[b]{0.485\columnwidth}
        \centering
        \includegraphics[width=\linewidth]{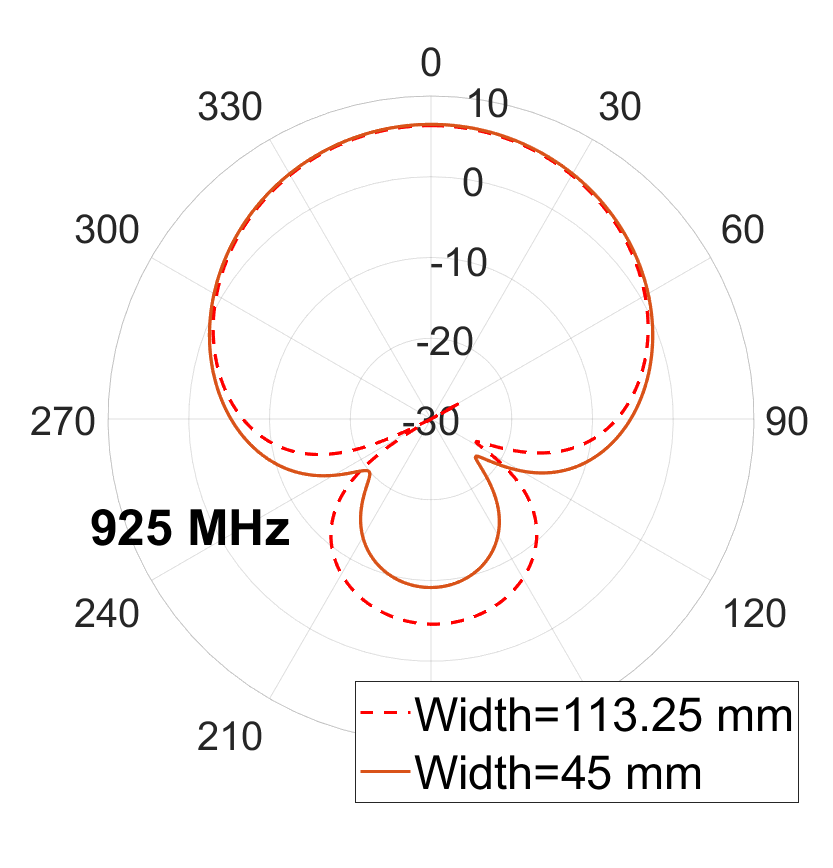}
        \caption{}
        \label{fig:polar_cut_925}
    \end{subfigure}
    \hfill
    \begin{subfigure}[b]{0.485\columnwidth}
        \includegraphics[width=\linewidth]{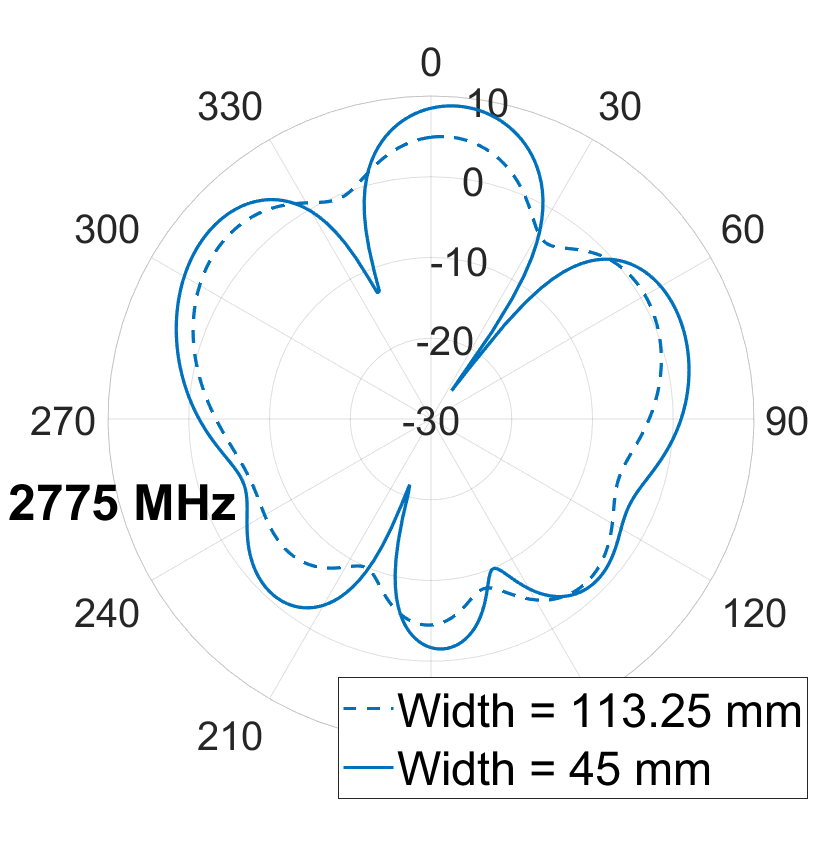}
        \caption{}
        \label{fig:polar_cut_2775}
    \end{subfigure}
    \begin{subfigure}[b]{0.48\columnwidth}
        \centering
        \includegraphics[width=\linewidth]{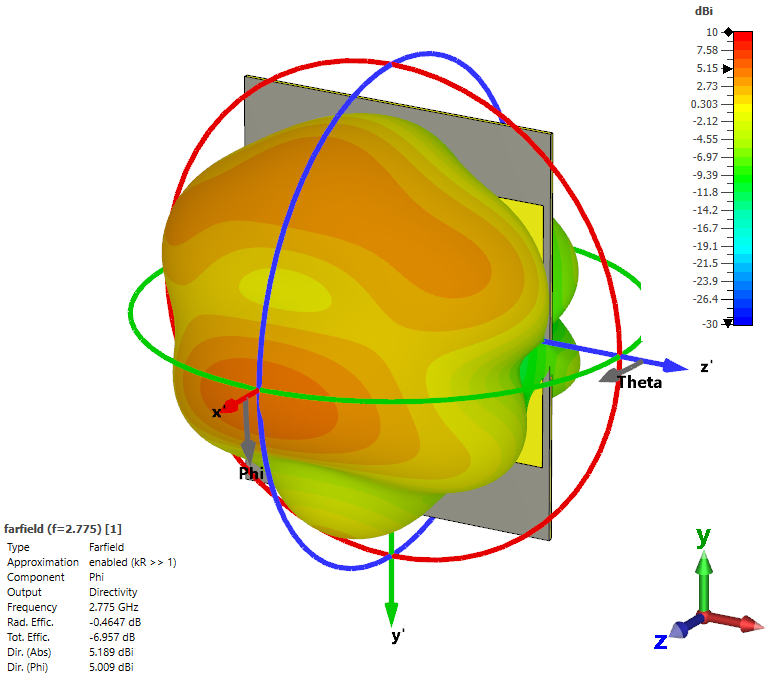}
        \caption{}
        \label{fig:3d_directivity_2775_w=113.25mm}
    \end{subfigure}
    \hfill
    % --- Right image ---
    \begin{subfigure}[b]{0.48\columnwidth}
        \centering
        \includegraphics[width=\linewidth]{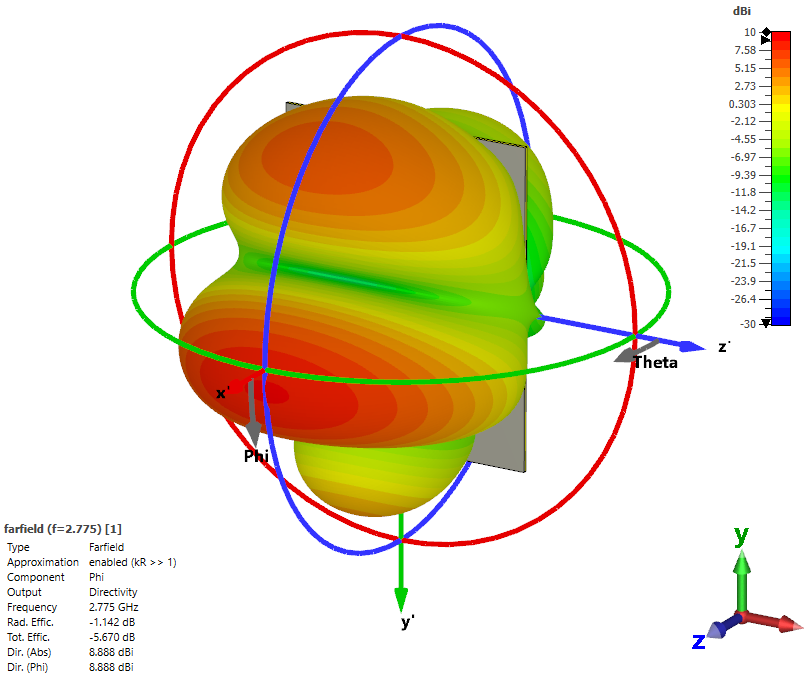}
        \caption{}
        \label{fig:3d_directivity_2775_w=45mm}
    \end{subfigure}
    \caption{(a) RPA width reduction. (b) Maintained longitudinal modes. (c) Directivity pattern in the fundamental frequency of 925 MHz in the XY plane in simulation. (d) Directivity pattern in the third harmonic frequency of 2775 MHz in the XY plane in simulation. Results are shown for patch width of 113.25mm and 45mm. 3D Directivity pattern in the third harmonic when the patch width is (e) 113.25mm and (f) after reducing the width to 45mm. }
    \label{fig:clipper_pcb_and_impedance}
\end{figure}

Reducing the width suppresses the transverse TM\textsubscript{mn} modes that exist in the wider configuration while preserving the longitudinal modes. Figure~\ref{fig:rpa-modes} presents a modal analysis performed in CST Studio, where the Q-factor and mode distribution are plotted as a function of frequency. The wide RPA (113.25 mm) supports 14 modes, whereas the narrow configuration (45 mm) supports only 7. Figure~\ref{fig:rpa_width_reduction} illustrates the transition to the narrower radiator, and Fig.~\ref{fig:rpa_maintained_modes} shows the E-field 
z-component distribution along the radiator for the fundamental and third harmonics. The results indicate that the dominant modes retained after width reduction are the TM\textsubscript{01} and TM\textsubscript{03} longitudinal modes.

The effect of width reduction is also evident in the radiation pattern. As shown in Fig.~\ref{fig:polar_cut_925}, the fundamental-frequency pattern remains nearly unchanged, whereas the third-harmonic pattern (Fig.~\ref{fig:polar_cut_2775}) becomes narrower in the XZ plane and exhibits deeper nulls, indicating reduced interference from transverse modes. This reduction increases peak directivity at the third harmonic by approximately 3.8 dB. Consequently, the optimized width produces an effective radiator with directive patterns at both the fundamental and third harmonics. Figures~\ref{fig:3d_directivity_2775_w=113.25mm} and~\ref{fig:3d_directivity_2775_w=45mm} show the simulated 3D radiation patterns before and after width reduction at 2775 MHz.

\section{Matching network}
 To obtain better 3rd harmonic conversion, the power delivered by the diodes to the RPA should be optimized. Therefore, a matching network is designed for it.
The matching network has two separate sections. First, The input is a 3-element pi section. This is designed to match the input impedance at the fundamental frequency of 925 MHz and is placed on the input microstrip feeding line. The pi-section is also a symmetrical low-pass filter that passes the fundamental frequency but blocks the 3rd harmonic from the input terminal.\\
The width reduction changed the input impedance from 15.7+j2.5$\Omega$ (-5.7dB return loss) to 82.4-j155.7$\Omega$ (-2.1dB return loss) at 925MHz and from 55.2-j22$\Omega$ (-13.5dB return loss) to 454.9+j136.5$\Omega$ (-1.75dB return loss) at 915MHz according to simulations.
 
\begin{figure}[!t]
    \centering
    % --- First image ---
    \begin{subfigure}{\linewidth}
        \centering
        \includegraphics[width=0.9\linewidth]{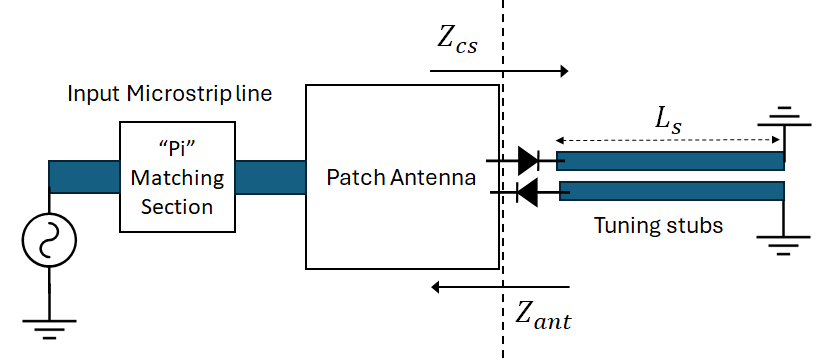}
        \caption{}
        \label{fig:match-rpa}
    \end{subfigure}

    \vspace{0.8em} % controls vertical spacing between images

    % --- Second image ---
    \begin{subfigure}{\linewidth}
        \centering
        \includegraphics[width=0.75\linewidth]{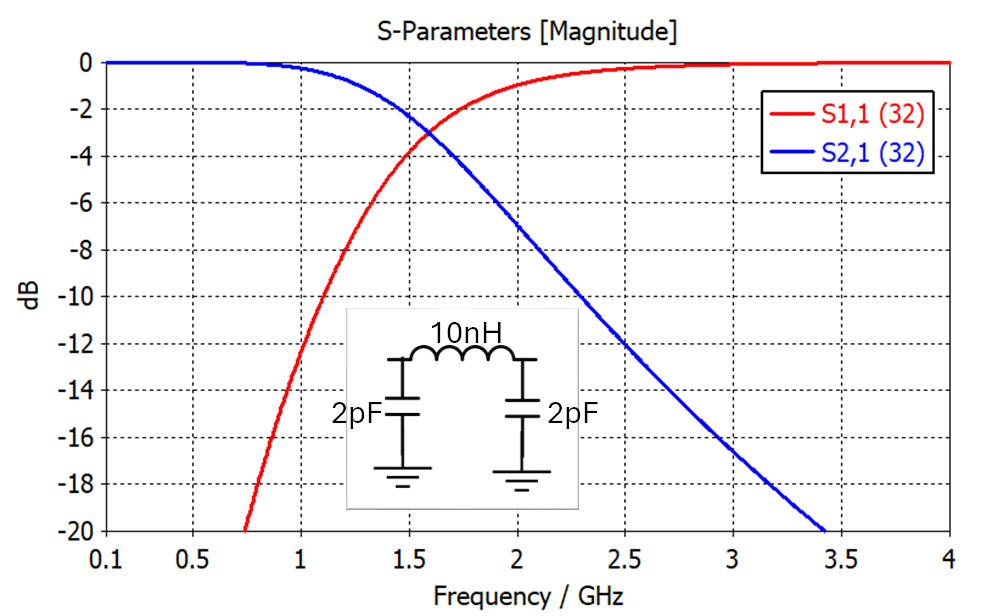}
        \caption{}
        \label{fig:lpf-pi}
    \end{subfigure}

    \caption{(a) Schematic description of the matching network. (b) Frequency domain response of the low pass filter at the input pi-section.}
    \label{fig:match-and-pi-lpf}
\end{figure}

 %As demonstrated in the following section, the results indicate that third-harmonic conversion efficiency improves significantly when this section is designed with an emphasis on low-pass filtering and third-harmonic suppression rather than impedance matching. 
The pi-section filter passes the fundamental frequency of 925MHz and blocks the 3rd harmonic at 2775MHz. It is made up of two 2pF capacitors and a 10nH inductor in a 'pi' circuit. The frequency response is depicted in Figure~\ref{fig:lpf-pi}.

The second part is connected to the clipper at the far edge of the RPA. Since the clipper acts as a secondary RF source to the RPA, its matching network is designed to optimize 3rd harmonic generation and radiation.
When the clipper sees a conjugate impedance at the third harmonic, this signal frequency component will be optimally generated and delivered back to the RPA. 
Each diode has a shorted microstrip line connected to its terminal. The shorted line acts as a tuning stub in series with the diode and shunt to the patch.
When varying the stub length, an increase in the 3rd harmonic conversion ratio is obtained.  Radiation efficiency in the first harmonic remains unchanged.\\

We will define two parameters: Z\textsubscript{ant}  representing the impedance seen when looking left to the RPA and input section, and Z\textsubscript{cs} which is the impedance seen when looking right to the clipper and stubs section as shown in Figure~\ref{fig:match-rpa}.
A passive simulation of the antenna and the PI section yielded the values Z\textsubscript{ant}=58.3+j17$\Omega$ at 925MHz and 154.1+j9.7$\Omega$ at f=2775MHz.
As mentioned in~\cite{Adamski}, to maximize harmonic conversion efficiency, the clipper impedance should ideally be the complex conjugate of the antenna. Therefore, the design goal is to obtain Z\textsubscript{cs}=Z\textsubscript{ant}* and to minimize the reflections in the third harmonic.
In the simulations, the swept parameters were P\textsubscript{in}, the RF input power at the fundamental frequency, and L\textsubscript{s}, the length of the stubs. 
The clipper and stub sections are power dependent. Its impedance values were obtained from a time-domain simulation. The optimal results were obtained for L\textsubscript{s}=24mm and for input power of P\textsubscript{in}=-3dBm at 925MHz.

Simulations showed that the value of Z\textsubscript{cs}=47.7+j55$\Omega$, minimized reflections and produced the strongest conversion to the 3rd harmonic (although it is not the complex conjugate of Z\textsubscript{ant}).
It was also noted that the shorted stubs have an intrinsic low radiation efficiency at 2775MHz, along with the main patch radiator.

\begin{figure}[!t]
    \centering
    % --- First image ---
    \begin{subfigure}{\linewidth}
        \centering
        \includegraphics[width=0.97\linewidth]{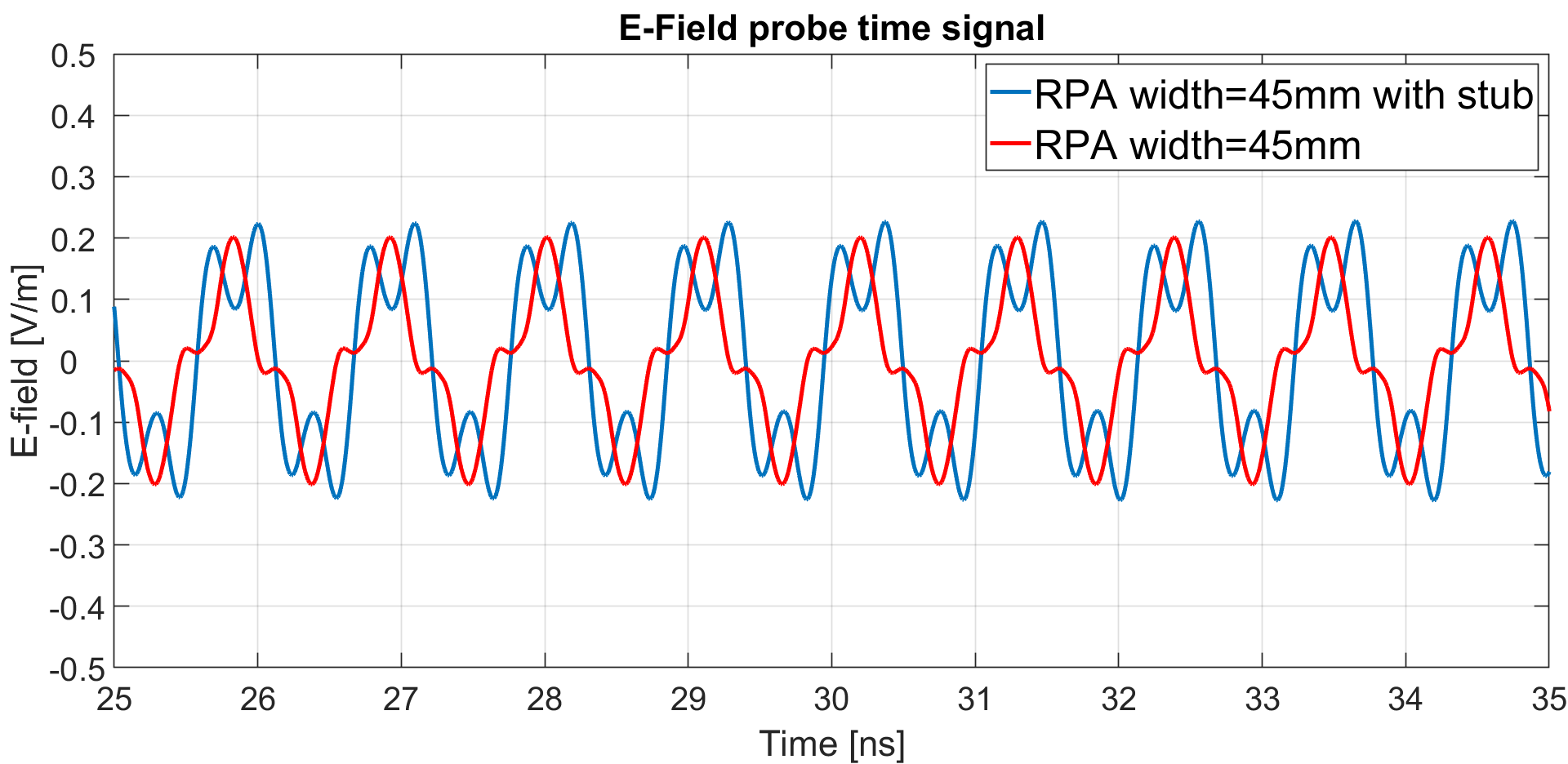}
        \caption{}
        \label{fig:match-rpa-stub-time}
    \end{subfigure}

    \vspace{0.8em} % controls vertical spacing between images

    % --- Second image ---
    \begin{subfigure}{\linewidth}
        \centering
        \includegraphics[width=0.97\linewidth]{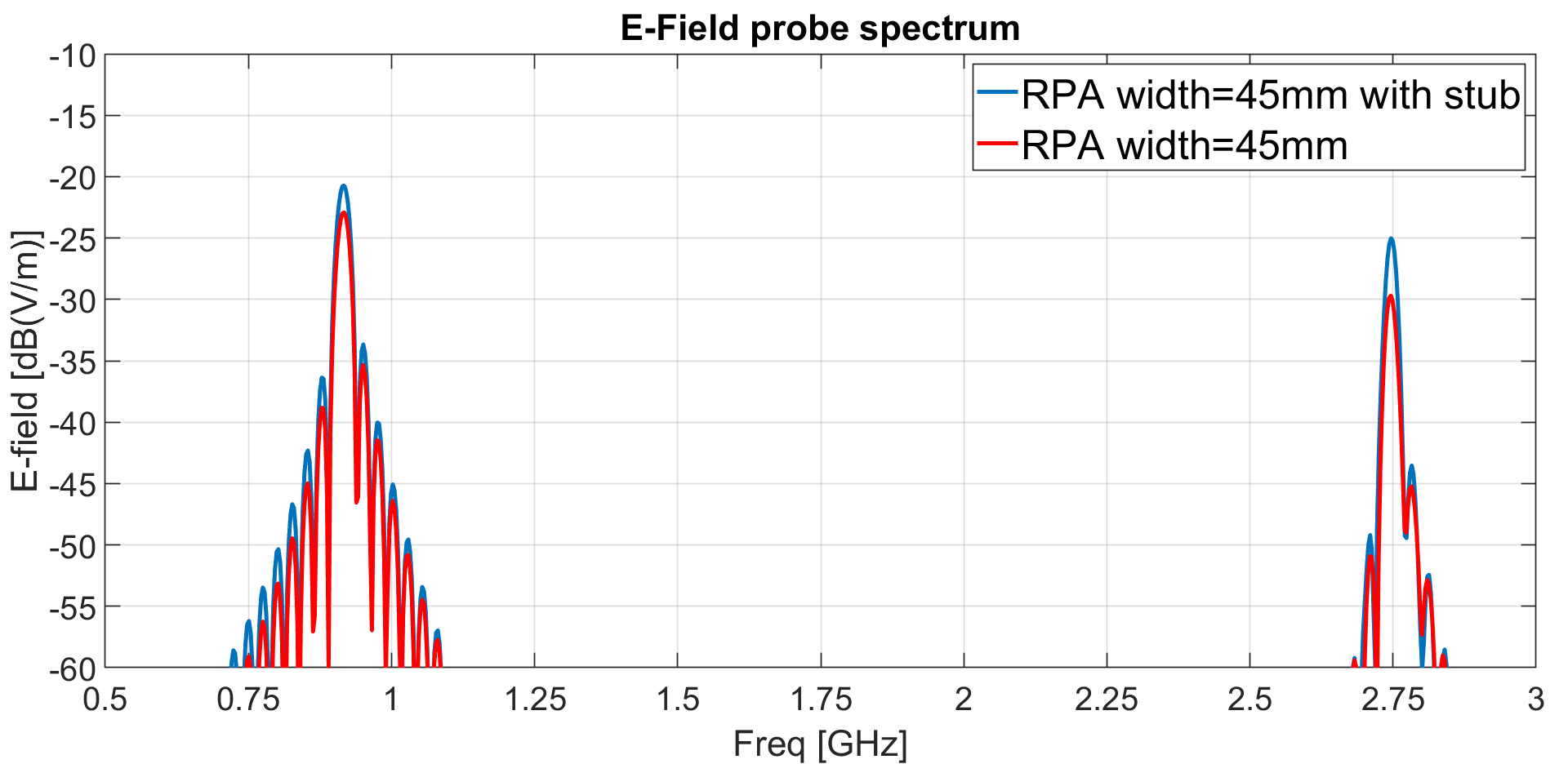}
        \caption{}
        \label{fig:match-rpa-stub-freq}
    \end{subfigure}

    \caption{(a) Time domain signal recorded in the E-field probe before and after adding the matching stub. (b) Frequency domain spectrum of the E-field probe before and after addition of the matching stub. }
    \label{fig:match_stub_time_freq_effect}
\end{figure}

The improvement of the third-harmonic efficiency by the matching network is clearly noticed in both the frequency spectrum and the time domain. Zooming in on the probe signal, one can see that it resembles a square wave. This indicates the presence of high-amplitude odd harmonics, see Figure~\ref{fig:match_stub_time_freq_effect}.
The highest conversion ratio obtained in the simulations was the 4.2dB difference between the 1st and 3rd harmonics.

\section{Measurements}
\subsection{Proposed antenna design and prototyping}
\begin{figure}[!t]
    \centering
    % --- First image ---
    \begin{subfigure}{\linewidth}
        \centering
        \includegraphics[width=1.1\linewidth]{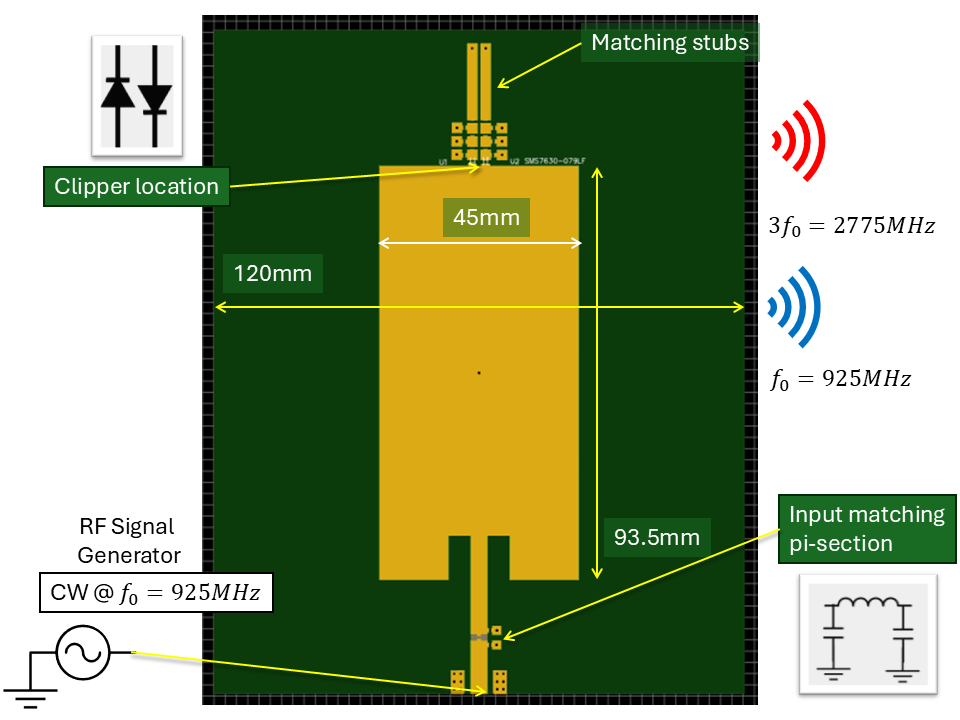}
        \caption{}
        \label{fig:pcb_top}
    \end{subfigure}

    \vspace{0.8em} % controls vertical spacing between images

    % --- Second image ---
    \begin{subfigure}{\linewidth}
        \centering
        \includegraphics[width=0.9\linewidth]{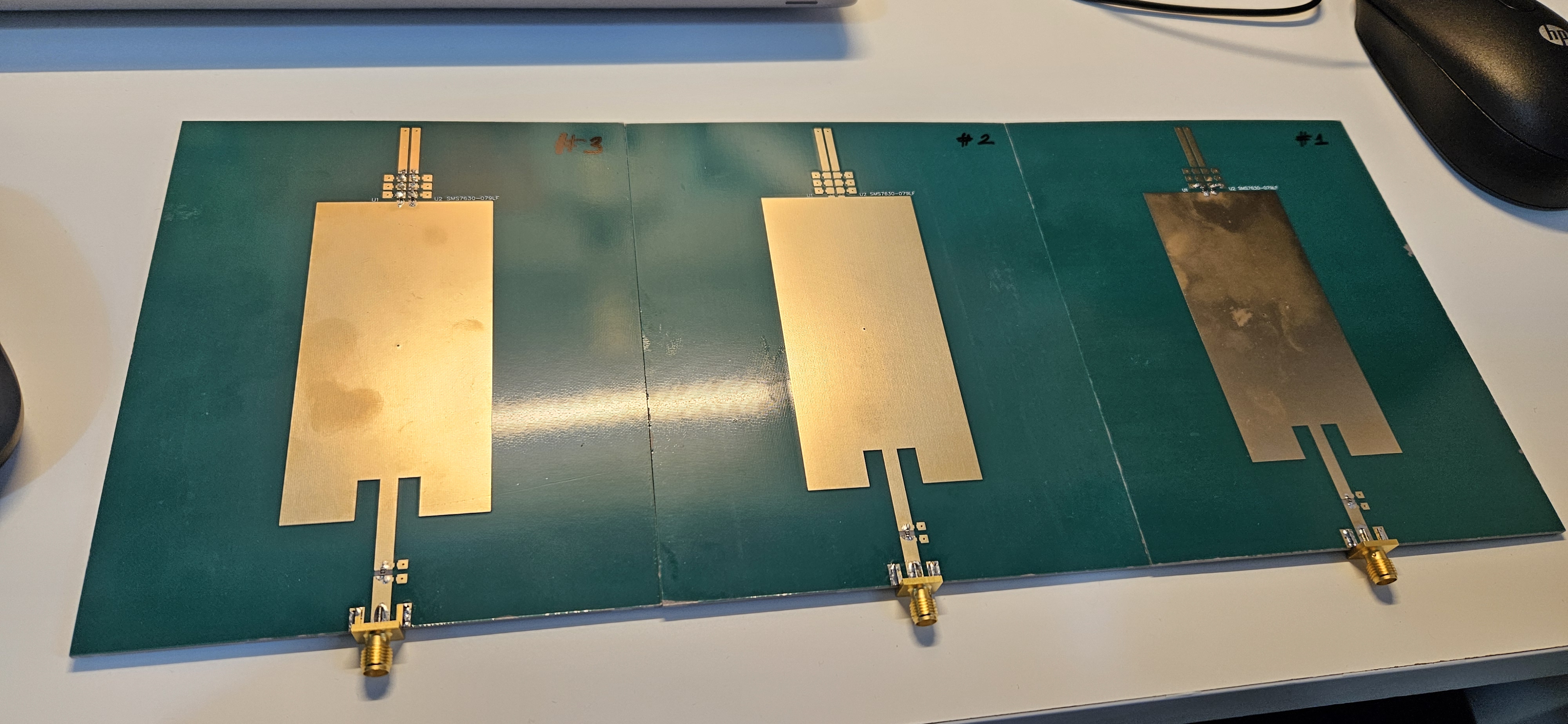}
        \caption{}
        \label{fig:three_fabricated_samples}
    \end{subfigure}

    \caption{(a) Proposed antenna design scheme and its dimensions. In the active measurement setup, a RF signal generator excites the antenna at the fundamental frequency f\textsubscript{0}. As a result, spontaneous radiation occurs in both fundamental frequency, f\textsubscript{0}, and the third harmonic, 3f\textsubscript{0}. (b) Fabricated RPA samples.  }
    \label{fig:pcb_top_and_samples}
\end{figure}

To validate our work so far, we have fabricated several prototypes of the simulated the antenna model, including the input pi-section and the top matching stubs. Figure~\ref{fig:pcb_top} is a top-layer view of the antenna in the PCB editor. The dielectric substrate is ZYF300CA-P, a 60mil Teflon-based material. 
Its reported parameters are $\epsilon_r$=3,and $\tan\delta$=0.0018 at 10GHz. These substrate losses are significantly lower than FR-4 and has produced near 80\% radiation efficiency in the simulations.
The clipper is a set of two back-to-back SMS7630 Schottky diodes in a SC-79 package.
The RPA has a 50$\Omega$ microstrip line feed and an inset perturbation.
Shorting via is added at the center of the patch for additional suppression of even modes and DC grounding. The input pi-section, tuning stubs, and optional matching pads can host 0603 or 0402 component sizes. 
Figure~\ref{fig:three_fabricated_samples} shows the manufactured antenna samples with SMA end-launch connectors on the input microstrip line.

\subsection{Passive measurements}
\begin{figure}[!t]
    \centering
    % --- First image ---
    \begin{subfigure}{\linewidth}
        \centering
        \includegraphics[width=0.6\linewidth]{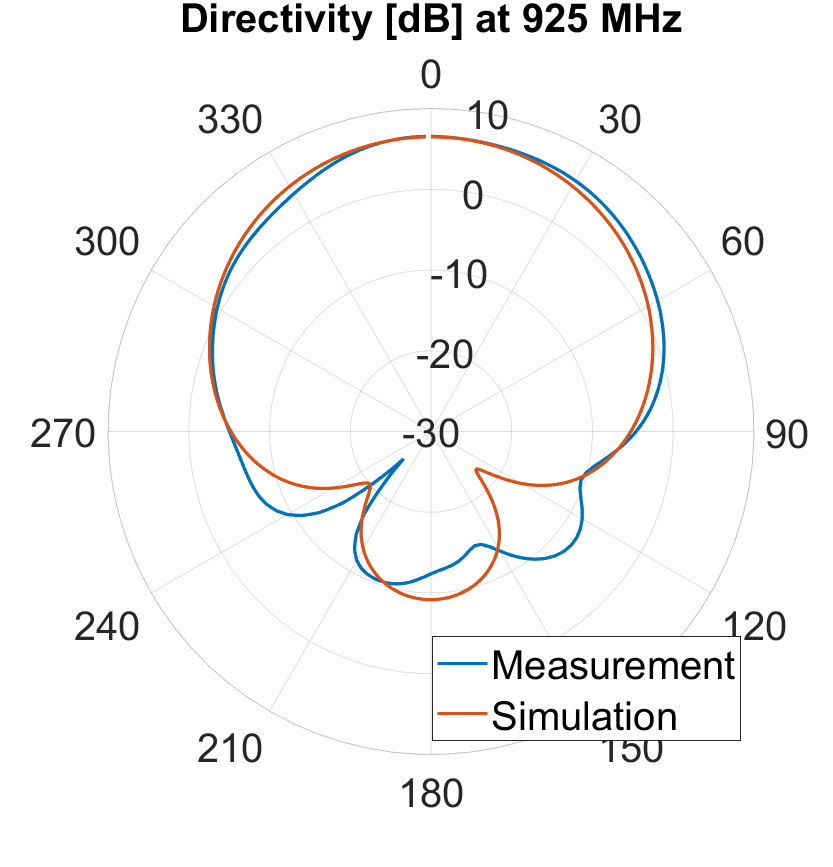}
        \caption{}
        \label{fig:RPA_passive_polar_patterns_925}
    \end{subfigure}

    \vspace{0.8em} % controls vertical spacing between images

    % --- Second image ---
    \begin{subfigure}{\linewidth}
        \centering
        \includegraphics[width=0.6\linewidth]{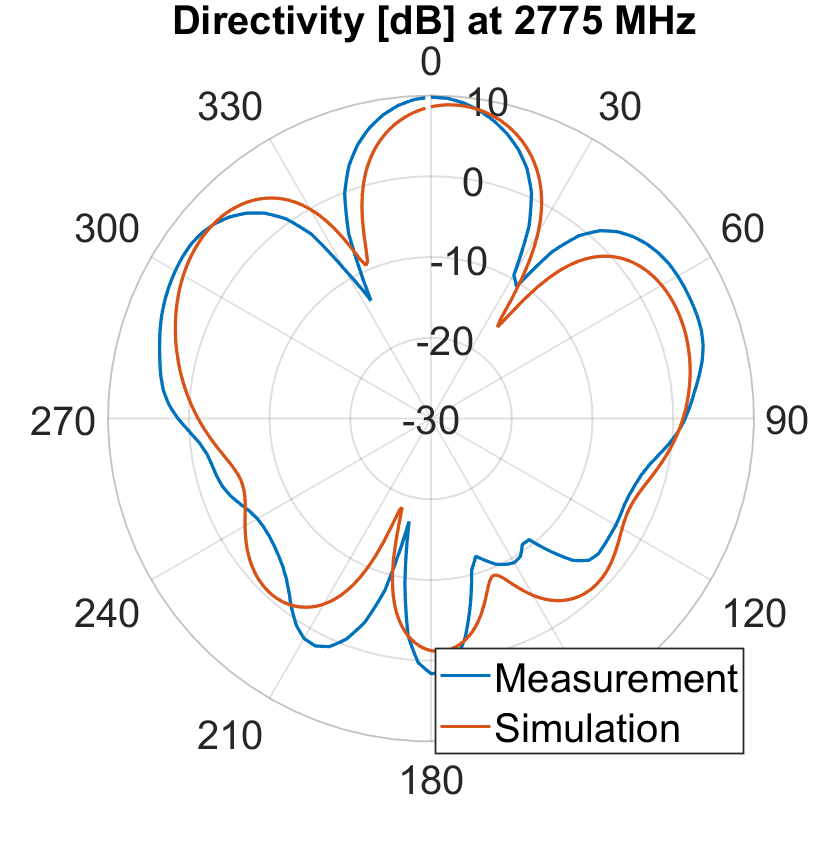}
        \caption{}
        \label{fig:RPA_passive_polar_patterns_2775}
    \end{subfigure}

\vspace{0.8em} % controls vertical spacing between images

    % --- Third image ---
    \begin{subfigure}{\linewidth}
        \centering
        \includegraphics[width=0.9\linewidth]{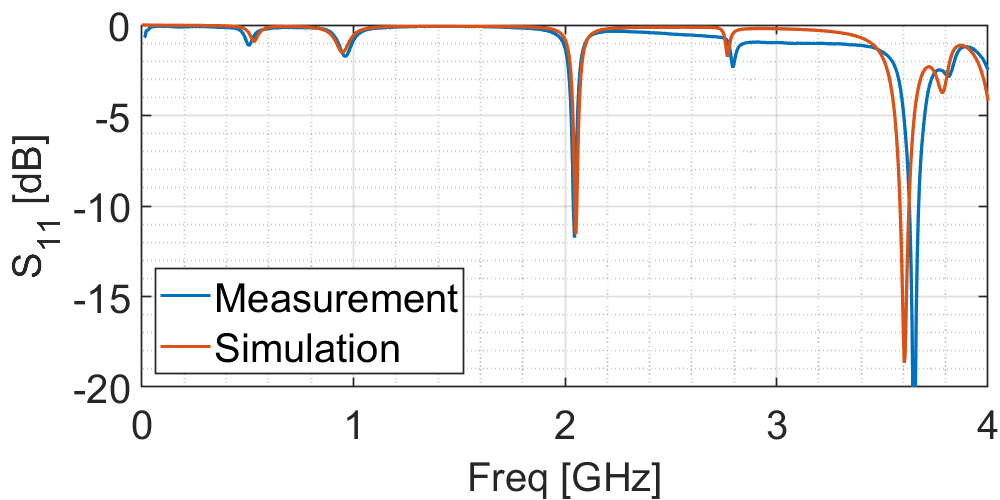}
        \caption{}
        \label{fig:SP_comp}
    \end{subfigure}
    
    \caption{Simulated directivity elevation patterns of the passive RPA in 925 MHz and (b) in 2775 MHz, compared to their corresponding prototype measurements. (c) S-parameters of the passive RPA. Simulations vs. Measurements.}   
    \label{fig:pcb_top_and_samples}
\end{figure}

In this passive measurement setup, a series 0$\Omega$ 0603 SMT package resistor is used to short the input pi-section. The diode clipper part is not mounted at this stage. The RPA was measured with a VNA and in an anechoic chamber to validate its basic frequency response and radiation patterns.
The radiation patterns and the S-parameters agree very well with the simulation results. They are presented in figures~\ref{fig:RPA_passive_polar_patterns_925}, ~\ref{fig:RPA_passive_polar_patterns_2775} and~\ref{fig:SP_comp}.

\subsection{Active measurements}
In this part, we mount the Schottky diode clipper on the RPA and investigate its RF power-dependent behavior. \\
The mounting setup on the RPA PCB is the one that produced the optimal 3\textsuperscript{rd} harmonic conversion results. It includes the pi-section low pass filter, the Schottky diodes clipper and 0$\Omega$ SMT resistors that short the matching stubs in series with the clipper leads. A portable RF signal generator is connected to the antenna connector to measure harmonic radiation under self-biased RF CW conditions. Figure~\ref{fig:pcb_top} also includes a graphic representation of this setup.\\
Let us now define in equation~\ref{eq:conv_efficiency} the conversion efficiency of the n\textsuperscript{th} harmonic radiation. It is defined here as the
 maximum value of EIRP (Effective isotropic radiated power) at the n\textsuperscript{th} harmonic frequency divided by the input RF power at the fundamental frequency:

\begin{equation}
\eta_n = \frac{\text{Max(EIRP)}@n\textsuperscript{th} harmonic}{P_{\text{in}}@f_0}\label{eq:conv_efficiency}
\end{equation}

\begin{figure}[!t]
    \centering
    % --- First image ---
    \begin{subfigure}[b]{0.97\columnwidth}
        %\centering
        \includegraphics[width=\linewidth]{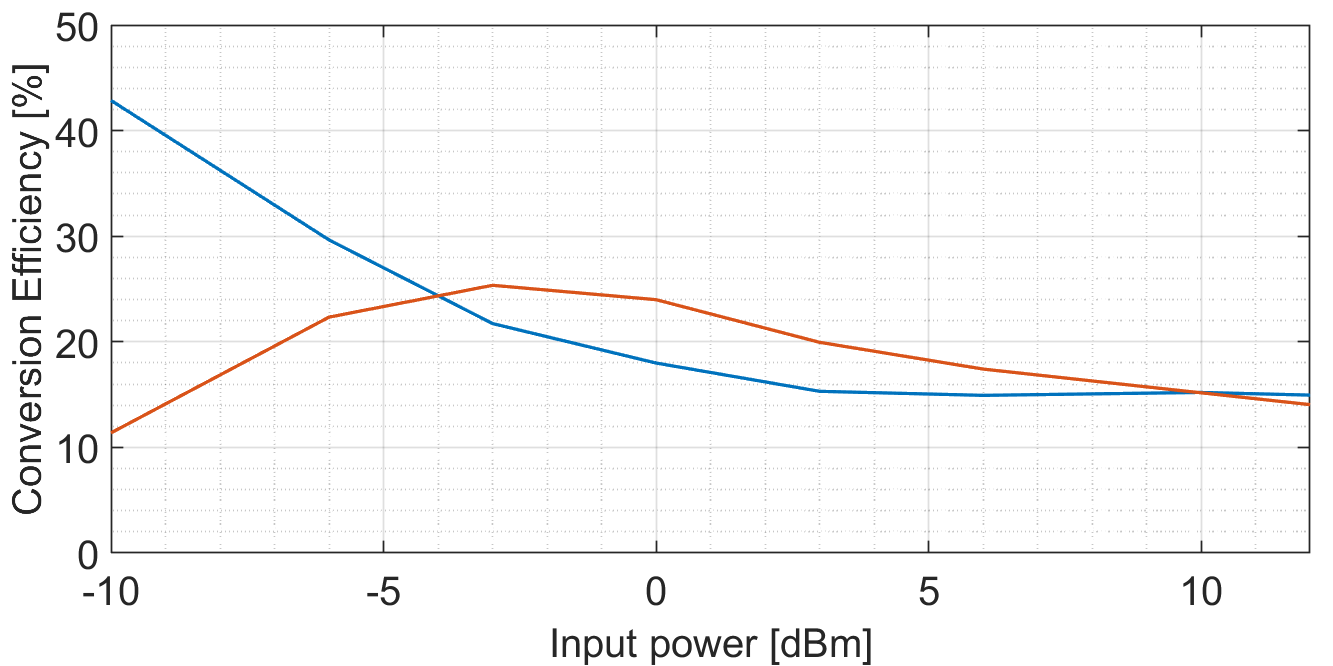}
        \caption{}
        \label{fig:harmonic_conv_eff}
    \end{subfigure}

\vspace{0.8em} % controls vertical spacing between images

    % --- Second image ---
    \begin{subfigure}[b]{\columnwidth}
        \centering
        \includegraphics[width=1\linewidth]{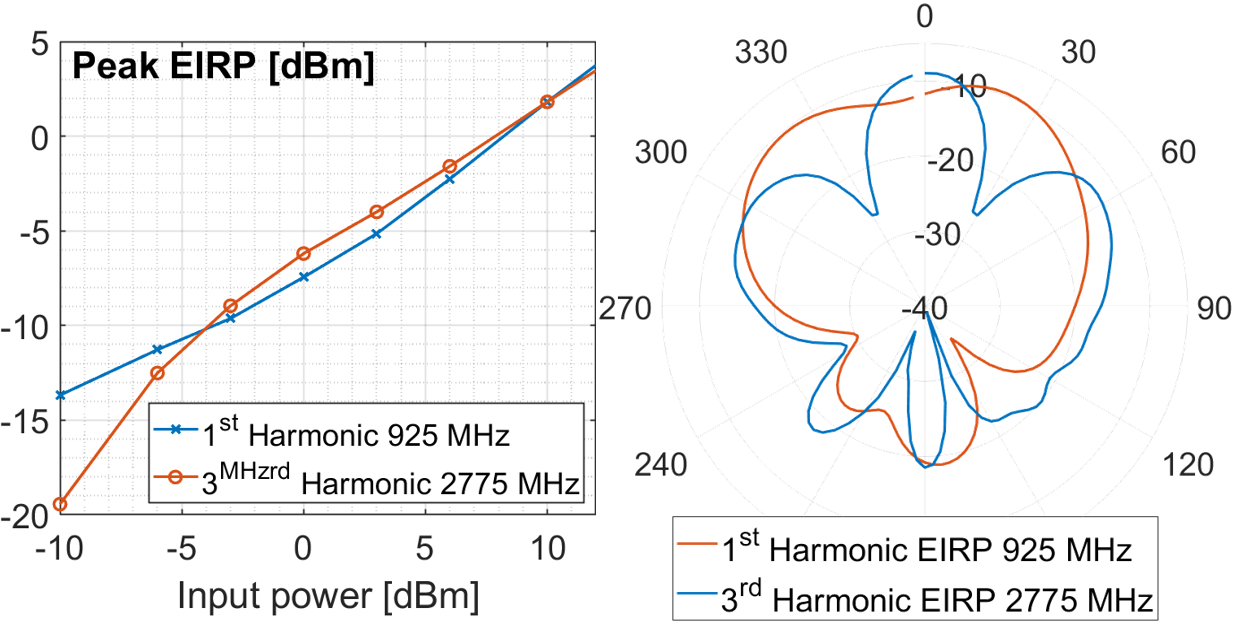}
        \caption{}
        \label{fig:polar_eirp_peak_eirp}
    \end{subfigure}
    
    \caption{(a) Conversion efficiency in the 1\textsuperscript{st} and the 3\textsuperscript{rd} harmonics. (b) Measured peak EIRP values (left) and measured radiation patterns in 925MHz and 2775MHz for RF input power of -3 dBM at the fundamental frequency of 925 MHz. The values represent measured EIRP in dBm. }   
    \label{fig:pcb_top_and_samples}
\end{figure}

Inside the anechoic chamber, the assembled RPA prototype is injected with a CW RF signal at 925 MHz. The input power ranges from -10 dBm to +13 dBm. Spontaneous radiation is emitted in both the fundamental and the third harmonic at 2775 MHz. The chamber measurement system produces complete radiation patterns in EIRP values. Since the RPA is a directive antenna, we record the maximal EIRP values and use them to calculate the third harmonic conversion efficiency. Figure~\ref{fig:polar_eirp_peak_eirp} (left) shows the peak EIRP values obtained in the measurement. Figure~\ref{fig:harmonic_conv_eff} plots the calculated conversion efficiency. An optimal value of 25\% was obtained at RF input power of -3dBm. The radiation pattern in EIRP values is plotted on the right side of Figure~\ref{fig:polar_eirp_peak_eirp} for a RF input power of -3dBm.

\subsection{Comparison to other published works}
Many works have been published which study passive harmonic tags in several application fields.The relevant ones are chosen, in order to compare them to the work presented in this article. We focus on passive, non-biased devices that generate higher order harmonics. The following Table~\ref{tab:conversion_comparison_with_notes} compares properties such  as: Output harmonic and frequency, device size, type of harmonic generator and conversion efficiency.

\begin{table*}[!t]
\centering
\begin{threeparttable}
\caption{Comparison of reported conversion efficiencies in prior published works}
\label{tab:conversion_comparison_with_notes}
\begin{tabular}{*{10}{>{\centering\arraybackslash}p{1.5cm}}}
\hline
\textbf{Work Reference} & \textbf{Pin [dBm]} & \textbf{Harmonic no.} & \textbf{Output frequency [GHz]} & \textbf{Harmonic generator} & \textbf{Antennas} & \textbf{Size [mm]} & \textbf{Electrical size in $\lambda$ of first harmonic} & \textbf{Conv. Efficiency [\%]} & \textbf{Application field} \\
\hline
~\cite{palazzi2015lowpower} & 0.0 & 2 & 2.08 & HSMS-2850 Single SBD\tnote{a} & Connectorized helix and patch antennas & 18x19 \tnote{b} & 0.06x0.066\tnote{b} & 15.524 & RFID \\
~\cite{gu2018improving} & -30.0 & 2 & 7.0 & SMS7630-079LF Single SBD & - & 24x24 \tnote{b} & 0.28x0.28\tnote{b} & 1.585\tnote{c} & Radar \\
~\cite{presas2003high} & -15.53 & 2 & 2.6 & HSCH-9161 anti-parallel pair GaAs SBD & Two quarter wavelength shorted patch antennas & 28x32.2  & 0.12x0.14 & 4.266 & RFID/Sensor \\
~\cite{colpitts1998harmonic} & -0.5 & 2 & 18.82 & HSCH-5340 SBD & Wire dipole & 12x1  & 0.38x0.03 & 0.04 & Radar \\
~\cite{kumar2020harmonic} & 8.0 & 3 & 2.778 & Commercial RFID Chip, Higgs 3 SOT & Two meandering dipole antennas & 46x30  & 0.142x0.09 & 0.00027 & RFID \\
~\cite{chow2006integrated} & 13.0 & 3 & 435.0 & HBV NLTL\tnote{d} 6 Sections & WR-22 waveguide section & 1.3x0.65x5 \tnote{e} & 0.63x0.31x2.42 & 1.5 & Submillimeter \& Terahertz radiators \\
~\cite{hollung2000distributed} & 22.0 & 3 & 130.5 & HBV NLTL 15 Sections & WR-5 waveguide section & 5.7x2.8x15  & 0.83x0.4x2.2 & 7.079 & solid state millimeter wave source \\
\textbf{This work} & \textbf{-3.0} & \textbf{3} & \textbf{2.775} & \textbf{SMS7630-079LF Anti-parallel pair} & \textbf{Single patch antenna} & \textbf{150x120 } & 0.46x0.37 &\textbf{25.35} & \textbf{Wireless communication} \\
\hline
\end{tabular}
\begin{tablenotes}
\footnotesize
\item[a] SBD=Schottky Barrier Diode.
%\item[b] Conversion efficiency was calculated from data provided in the article.
\item[b] Size does not include commercial connectorized antennas used in the said work.
\item[c] Only 2-port conducted conversion efficiency was reported. Over the air efficiency values are not clearly stated.
\item[d] HBV=Heterostructure Barrier Varactors.
\item[e] Size is estimated from the featured images in this article.
\end{tablenotes}
\end{threeparttable}
\end{table*}

\section{Conclusion}

This work has demonstrated that a rectangular patch antenna (RPA) incorporating a self-biased Schottky-diode clipper can efficiently generate and radiate odd harmonics without any external biasing. Embedding the nonlinear load at high-field regions, together with tailored antenna geometry and dual matching networks, enabled strong third-harmonic excitation and radiation. In particular, reducing the antenna width suppressed unwanted transverse modes and yielded a directive radiation pattern at the third harmonic. The fabricated prototypes achieved a measured peak third-harmonic radiated conversion efficiency of 25.35\% at an input power of –3dBm.

The optimized configuration produced third-harmonic EIRP that exceeded the radiated power at the fundamental frequency under optimal conditions, confirming the effectiveness of passive, geometry-driven harmonic enhancement in compact antennas. Within the class of passive, unbiased diode-loaded radiators, the proposed design exhibits one of the highest reported third-harmonic radiation efficiencies.

These results highlight the potential of integrated nonlinear loading as a compact mechanism for passive harmonic control, enabling frequency-agile, spectrum-efficient, and multifunctional wireless systems without the need for active circuitry or auxiliary bias networks.

%The RPA spontaneously radiates significant power in the 1\textsuperscript{st} and 3\textsuperscript{rd} harmonics.
%The RPA efficiently converts 1\textsuperscript{st} to 3\textsuperscript{rd} harmonics.
%The Matched RPA increases the power conversion to the 3\textsuperscript{rd} harmonic and enhances the effect. The peak EIRP of the 3\textsuperscript{rd} harmonic is higher than the 1\textsuperscript{st} harmonic when the conversion is optimal.
%Optimal conversion efficiency that we obtained is 25\% when the RF input power in the fundamental frequency is -3 dBm.

% if have a single appendix:
%\appendix[Proof of the Zonklar Equations]
% or
%\appendix  % for no appendix heading
% do not use \section anymore after \appendix, only \section*
% is possibly needed

% use appendices with more than one appendix
% then use \section to start each appendix
% you must declare a \section before using any
% \subsection or using \label (\appendices by itself
% starts a section numbered zero.)
%

\appendices
\section{Frequency domain analysis of the clipper circuit}
\label{appendix: Clipper_freq_domain_analysis}

As explained in~\cite{Adamski}, we can also describe the steady-state periodic response of the clipper circuit in Figure~\ref{fig:ClipperSchematicMain} in terms of a single sided complex Fourier series.
The total current is:
\begin{equation}
I_{total}(t) = {\rm Re}\left\{\sum_{n=0}^{\infty} I(n)e^{jn\omega t}\right\}
\end{equation}
where $n$ is the harmonic index. Taking into account the source $V_s(t)$, its internal admittance $Y_{src}$, and the diode branches, we can write:
\begin{equation}
I_{total}(t) = {\rm Re}\left\{\sum_{n=0}^{\infty}[Y_{src}(n)V_s(n)+i_1(n)+i_2(n)]e^{jn\omega t}\right\}. \label{eq:SumtotalCurrent}
\end{equation}

The diode branches are identical except for the opposite directions of the current flow. Each branch corresponds to half a period of the sinusoidal excitation waveform according to its polarization. Therefore, the relation between the branch currents is as follows:
\begin{align}
   i_2(n)=-i_1e^{-jn\pi}
\end{align}
The odd harmonics will add up:
\begin{align}
   i_2(2n+1)=i_1(2n+1), \quad{ n=0,1,...} \label{eq:oddTerm}
\end{align}
And the even harmonics and DC term will cancel out:
\begin{align}
   i_2(2n)=-i_1(2n),\quad{ n=0,1,...} \label{eq:evenTerm}
\end{align}
Plugging~\eqref{eq:oddTerm} and~\eqref{eq:evenTerm} into expression~\eqref{eq:SumtotalCurrent}:
\begin{align}
   I_{total}(t) &= \mbox{Re}\left\{[Y_{src}(1)V_s(1)+2i_1(1)]e^{j\omega t}+\right. \notag\\
   &\quad +\left.[2i_1(3)]e^{j3\omega t}+[2i_1(5)]e^{j5\omega t}+...\right\} \label{eq:SumOddTermsOnly}
\end{align}
The last expression demonstrates how eventually the total current is composed of fundamental and odd harmonics only and the DC and even terms are canceled. Figure~\ref{fig:FFTResults} shows the resulting odd-harmonic spectrum of the total current in the simulations of our analytical model.

\section{Validation of the simulations approach}
\label{appendix: Validation_smv1405}

%Appendix one text goes here.

To ensure the accuracy of our antenna simulations, we validated the results against experimental measurements. Several diodes were tested on a custom-fabricated PCB coupon made from a 60 mil FR-4 substrate, where the diodes were connected in shunt to the transmission line, as shown in Fig.~\ref{fig:smv1405-sp}. At this stage, a varactor diode was used for the evaluation.

The diode response was characterized under both small-signal and large-signal excitation conditions.

\subsection{Small signal results}
The frequency response of the PCB coupon was measured using a vector network analyzer (VNA) and compared with a broadband S-parameter simulation. The VNA output power was set to -15 dBm
 to ensure small-signal operation, thereby validating the linear-passive behavior of the diode. Excellent agreement was achieved between the simulated and measured results, confirming the accuracy of the model under small-signal conditions. Figure~\ref{fig:smv1405-sp} shows the results for the SMV1405 varactor diode in an SC-79 package from Skyworks~\cite{SMV1405}.

\subsection{Large signal results}
A CW signal with varying amplitude was injected into the PCB coupon using an RF signal generator. The 1\textsuperscript{st}, 2\textsuperscript{nd}, and 3\textsuperscript{rd} harmonics were measured with a spectrum analyzer.

The corresponding simulation replicates this setup by exciting the circuit with a single-frequency source of varying input power, while harmonic extraction is performed through FFT analysis during post-processing. This procedure validates the nonlinear behavior of the diode.

The excitation tone was set to f\textsubscript{0}=1 GHz, and the input power was swept from -15 to +17dBm. The output harmonic powers were recorded at each level and normalized to the fundamental component to enable comparison of the relative conversion ratios.

\begin{figure}[!t]
    \centering
    % --- First image ---
    \begin{subfigure}{\linewidth}
        \centering
        \includegraphics[width=0.9\linewidth]{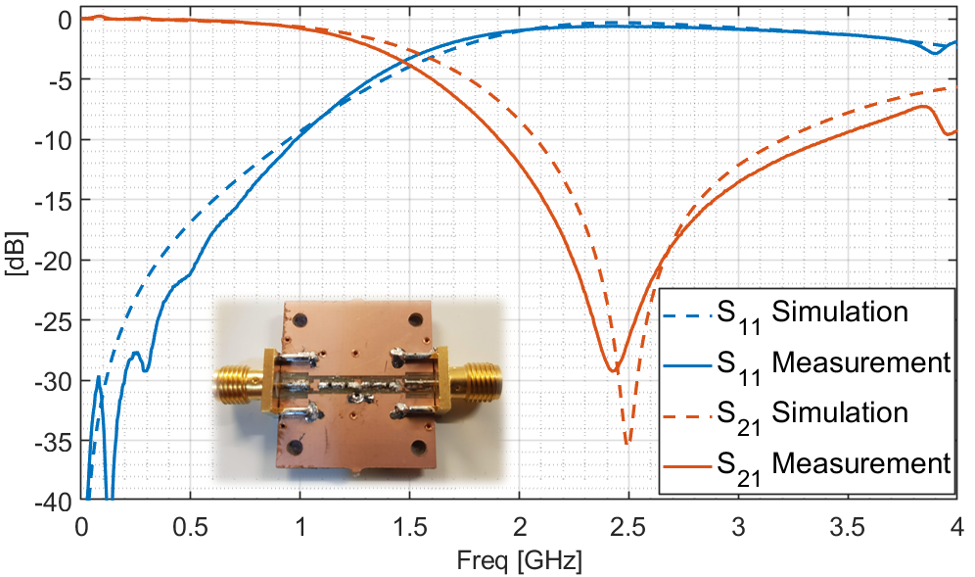}
        \caption{}
        \label{fig:smv1405-sp}
    \end{subfigure}

    \vspace{0.8em} % controls vertical spacing between images

    % --- Second image ---
    \begin{subfigure}{\linewidth}
        \centering
        \includegraphics[width=0.7\linewidth]{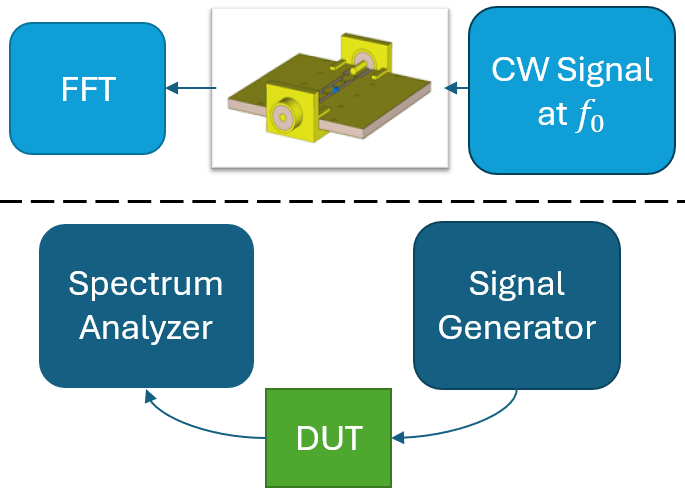}
        \caption{}
        \label{fig:large_signal_setups}
    \end{subfigure}

\vspace{0.8em} % controls vertical spacing between images

    % --- Third image ---
    \begin{subfigure}{\columnwidth}
        \centering
        \includegraphics[width=\columnwidth]{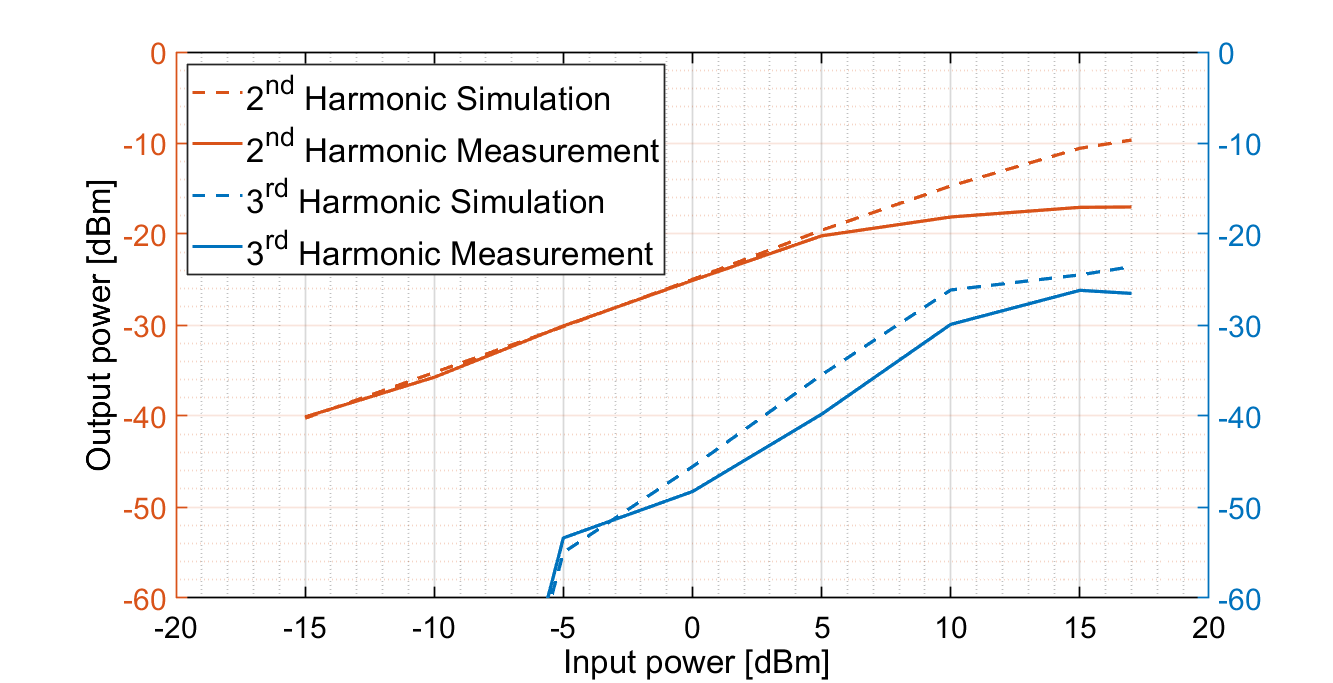}
        \caption{}
        \label{fig:smv1405_harmonic_large_signal}
    \end{subfigure}

    \caption{(a) A Comparison between simulated and measured S-Parameters of a SMV1405 varactor diode. The diode is connected in shunt to the microstrip line. The VNA output power is -15dBm.(b) Block diagrams of the simulation setup (top) and measurement setup (bottom). A single tone is injected at f\textsubscript{0}=1 GHz. The input power varies between -15 and +17 dBm. (c) Second and third harmonic measured output power compared to the simulation. f\textsubscript{0}=1 GHz. }   
    \label{fig:pcb_top_and_samples}
\end{figure}

Figure~\ref{fig:large_signal_setups} presents the block diagrams of both the simulation and measurement configurations. Figure~\ref{fig:smv1405_harmonic_large_signal} shows the comparison between the simulated and measured responses of the SMV1405 varactor diode in the second and third harmonics.

\subsection{Observations}
Several varactor and PIN diodes were characterized through both simulation and measurement. Components that included vendor-supplied SPICE models showed strong correlation between the simulated and measured second- and third-harmonic responses. These results demonstrate that SPICE models derived from empirical device data provide a reliable and accurate basis for nonlinear harmonic prediction.
% you can choose not to have a title for an appendix
% if you want by leaving the argument blank
%\section{}
%Appendix two text goes here.

% use section* for acknowledgment
\section*{Acknowledgment}

The authors would like to thank Mr. Itzhak Shapir for useful discussions and conversations.

% Can use something like this to put references on a page
% by themselves when using endfloat and the captionsoff option.
\ifCLASSOPTIONcaptionsoff
  \newpage
\fi

% trigger a \newpage just before the given reference
% number - used to balance the columns on the last page
% adjust value as needed - may need to be readjusted if
% the document is modified later
%\IEEEtriggeratref{8}
% The "triggered" command can be changed if desired:
%\IEEEtriggercmd{\enlargethispage{-5in}}

% references section

% can use a bibliography generated by BibTeX as a .bbl file
% BibTeX documentation can be easily obtained at:
% http://mirror.ctan.org/biblio/bibtex/contrib/doc/
% The IEEEtran BibTeX style support page is at:
% http://www.michaelshell.org/tex/ieeetran/bibtex/
%\bibliographystyle{IEEEtran}
% argument is your BibTeX string definitions and bibliography database(s)
%\bibliography{IEEEabrv,../bib/paper}
%
% <OR> manually copy in the resultant .bbl file
% set second argument of \begin to the number of references
% (used to reserve space for the reference number labels box)

% Added comment to force bibliography update
\bibliographystyle{IEEEtran} % IEEE style
\bibliography{references}    % references.bib file

% biography section
% 
% If you have an EPS/PDF photo (graphicx package needed) extra braces are
% needed around the contents of the optional argument to biography to prevent
% the LaTeX parser from getting confused when it sees the complicated
% \includegraphics command within an optional argument. (You could create
% your own custom macro containing the \includegraphics command to make things
% simpler here.)
%\begin{IEEEbiography}[{\includegraphics[width=1in,height=1.25in,clip,keepaspectratio]{mshell}}]{Michael Shell}
% or if you just want to reserve a space for a photo:

%\begin{IEEEbiography}{Yishai Brill}
%Biography text here.
%\end{IEEEbiography}

% if you will not have a photo at all:
%\begin{IEEEbiographynophoto}{Yakir Hadad}
%Biography text here.
%\end{IEEEbiographynophoto}

% insert where needed to balance the two columns on the last page with
% biographies
%\newpage

% You can push biographies down or up by placing
% a \vfill before or after them. The appropriate
% use of \vfill depends on what kind of text is
% on the last page and whether or not the columns
% are being equalized.

%\vfill

% Can be used to pull up biographies so that the bottom of the last one
% is flush with the other column.
%\enlargethispage{-5in}

% that's all folks
\end{document}